\documentclass[aps,prl,twocolumn,superscriptaddress]{revtex4}
\usepackage{graphicx}
\usepackage{latexsym}
\usepackage{amssymb}
\usepackage{amsmath}
\usepackage{amsfonts}
\usepackage{bm}
\usepackage{multirow}
\usepackage{color}
\newcommand{\bra}[1]{\langle #1|}
\newcommand{\ket}[1]{|#1 \rangle}
\newcommand{\dsZ}{\mathbb{Z}}
\newcommand{\dsR}{\mathbb{R}}
\newcommand{\dsC}{\mathbb{C}}

\newcommand{\Cl}{\mathcal{C\ell}}
\newcommand{\U}[1]{\mathrm{U}(#1)}
\newcommand{\SU}[1]{\mathrm{SU}(#1)}
\renewcommand{\O}[1]{\mathrm{O}(#1)}
\newcommand{\SO}[1]{\mathrm{SO}(#1)}
\newcommand{\ii}{\mathrm{i}}
\newcommand{\dd}{\mathrm{d}}

\renewcommand{\Re}{\mathop{\mathrm{Re}}}
\renewcommand{\Im}{\mathop{\mathrm{Im}}}
\newcommand{\vect}[1]{{\bm{#1}}}
\newcommand{\mat}[1]{\left[\begin{matrix}#1\end{matrix}\right]}
\newcommand{\eqnref}[1]{Eq.\,\eqref{#1}}
\newcommand{\figref}[1]{Fig.\,\ref{#1}}
\newcommand{\tabref}[1]{Tab.\,\ref{#1}}

\newcommand{\beq}{\begin{equation}}
\newcommand{\eeq}{\end{equation}}
\newcommand{\beqn}{\begin{eqnarray}}
\newcommand{\eeqn}{\end{eqnarray}}

\begin{document}

\title{Interacting Topological Insulator and Emergent Grand Unified Theory}

\author{Yi-Zhuang You}

\author{Cenke Xu}

\affiliation{Department of physics, University of California,
Santa Barbara, CA 93106, USA}

\begin{abstract}

Motivated by the Pati-Salam Grand Unified Theory~\cite{patisalam}
we study $(4+1)d$ topological insulators with $\SU4 \times\SU2_1
\times\SU2_2$ symmetry, whose $(3+1)d$ boundary has 16 flavors of
left-chiral fermions, which form representations $(\mathbf{4},
\mathbf{2}, \mathbf{1})$ and $(\bar{\mathbf{4}}, \mathbf{1},
\mathbf{2})$. The key result we obtain is that, without any
interaction, this topological insulator has a $\mathbb{Z}$
classification, namely any quadratic fermion mass operator at the
$(3+1)d $ boundary is prohibited by the symmetries listed above;
while under interaction this system becomes trivial, namely its
$(3+1)d$ boundary can be gapped out by a properly designed short
range interaction without generating nonzero vacuum expectation
value of any fermion bilinear mass, or in other words, its
$(3+1)d$ boundary can be driven into a ``strongly coupled
symmetric gapped (SCSG) phase". Based on this observation, we
propose that after coupling the system to a dynamical
$\SU4\times\SU2_1 \times\SU2_2$ lattice gauge field, the
Pati-Salam GUT can be fully regularized as the boundary states of
a $(4+1)d$ topological insulator with a {\it thin} fourth spatial
dimension, the thin fourth dimension makes the entire system
generically a $(3+1)d$ system. The mirror sector on the opposite
boundary will {\it not} interfere with the desired GUT, because
the mirror sector is driven to the SCSG phase by a carefully
designed interaction and is hence decoupled from the GUT.

\end{abstract}


\maketitle

\section{1. Introduction}

In the Standard Model of particle physics and the Grand Unified
Theories (GUT), the gauge coupling is asymmetric between left and
right handed fermions. This chiral gauge coupling makes it
difficult to regularize the field theory as a full quantum theory
on a lattice. The main obstacle of this lattice regularization is
the fermi doubling theorem~\cite{doublingA,doublingB}, which
states that both left and right handed fermions will arise at low
energy for any lattice fermion model. Then when the lattice
fermion is coupled to a gauge field, it will induce the same
coupling between left and right fermions, which is inconsistent
with the Standard Model or the GUT. In order to get around the
fermi doubling theorem, one method is to realize the GUT on the
$3d$ boundary of a $4d$ topological insulator (TI), or in other
words at the domain wall of the mass of $4d$ Dirac
fermion~\cite{domainwall,kaplan1992,kaplan2012}~\footnote{Throughout
the paper, $3d$ and $4d$ represent the spatial dimensions of the
boundary and bulk respectively, while $(3+1)d$ and $(4+1)d$
represent the space-time dimensions.}. Then there is a mirror
sector of fermions with opposite chirality localized on the other
opposite boundary, which is spatially separated from the GUT.
Fermions at each boundary can naturally have a chiral coupling to
the bulk gauge fields. However, this method requires subtle
adjustment of the scale of the fourth dimension: if the fourth
dimension is too large, the gauge boson in the bulk will be
gapless and interfere with the low energy physics of the boundary;
on the other hand if the fourth dimension is too small, then the
GUT suffers from interference with its mirror sector on the other
boundary~\cite{latticefermions}.

\begin{figure}[t]
\begin{center}
s\includegraphics[width=220pt]{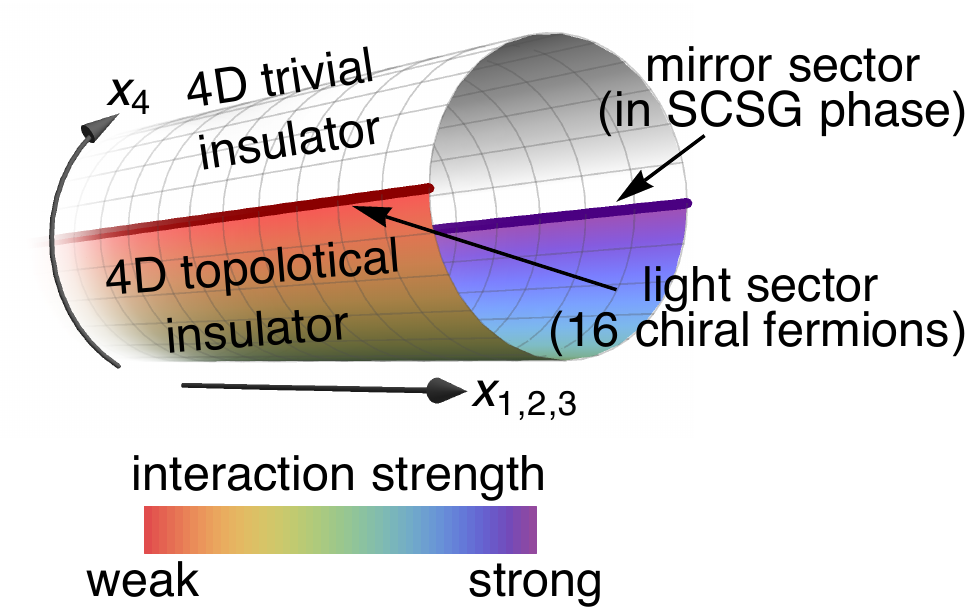} \caption{(Color
online.) Regularizing the GUT on a lattice with three extended
dimensions $x_{1,2,3}$ and a compactified dimension $x_4$. The
light sector (GUT) and the mirror sector are separated in the
$x_4$ dimension, as two $3d$ boundaries of a $4d$ TI. The mirror
sector is decoupled from GUT due to interaction, whose strength
varies with $x_4$.} \label{fig: 4dTI}
\end{center}
\end{figure}

In a GUT, effectively in every generation there are 16 left handed
fermions, thus its mirror sector must have 16 right handed
fermions with the same gauge coupling. It would be ideal if we can
gap out the mirror sector without affecting the fermions in the
GUT, $i.e.$ decouple the mirror sector from low energy physics
completely. Then we can regularize the GUT on the $3d$ boundary of
a $4d$ TI with a very thin fourth dimension (which makes the bulk
generically a $3d$ system), see \figref{fig: 4dTI}. However, if
the mirror sector is gapped out in the standard way, namely they
are gapped out by condensing a boson field that couples to the
mass operators of the mirror fermions, then the same boson field
would couple to the fermions in the GUT and gap them out as well.
Thus we seek the possibility to gap out the mirror sector while
having zero fermion bilinear expectation value, $\langle
\psi_a^\intercal \ii\sigma^y \psi_b\rangle = 0$ in the mirror
sector~\footnote{In condensed matter physics, $\psi^\intercal
i\sigma^y \psi$ is a Cooper pair operator; while in high energy
physics, it is the Majorana mass of chiral fermion.}, for
arbitrary flavor indices $a,b$ $= 1,...,16$. We label this fully
gapped phase of the mirror sector as ``strongly coupled symmetric
gapped phase" (SCSG phase).\footnote{In principle, it is also
possible to drive the mirror sector into a fully symmetric
topological order which has a gapped spectrum, but degenerate
ground states. This case was immensely studied in condensed matter
physics. But in our current work we focus on the case when the
mirror sector is nondegenerately gapped by interaction, $i.e.$
there is no topological order in the SCSG phase.} (This phase was
also called the ``strongly coupled symmetric phase" or the
``paramagnetic strong-coupling (PMS) phase'' in literature, see
appendix A for a review of the recent progress on the SCSG phase
in the condensed matter community.)

%We will now drop the primes for notational convenience, and it
%should be understood that when we attempt to gap out the fermions
%$\psi_a$ we intend to gap out the mirror sector while leaving the
%ordinary fermions gapless.

A SCSG phase with fully gapped but nondegenerate spectrum in the
mirror sector is only possible when the system satisfies the
following two necessary criteria:

{\it (1)} Based on the anomaly matching
condition~\cite{anomalymatching,anomalymatching2}, the system
should {\it not} have any symmetry that would be anomalous in the
mirror sector once the system is coupled to the gauge fields.

For instance the global charge $\U1$ symmetry of chiral fermions
$\psi_{a,L} \rightarrow e^{\ii\theta} \psi_{a,L}$, which was a key
assumption for the no-go theorem proved in
Ref.\,\cite{doublingA,doublingB}, should not exist in the lattice
model.

{\it (2)} The $4d$ bulk state is a nontrivial TI (hence must have
gapless chiral fermions on its $3d$ boundary) without interaction;
but under interaction it becomes a trivial state, which means that
its boundary can be driven into the desired SCSG phase by
interaction~\footnote{Based on the first criterion, the global
U(1) symmetry in the bulk must be explicitly broken, thus the term
``insulator" is not entirely accurate. We use the term topological
insulator because in our construction this anomalous U(1) symmetry
is only broken by the four fermion interaction term, while it is
preserved at the noninteracting level.}.

Notice that there must be a minimum nonzero critical interaction
strength for this SCSG phase to exist, because a weak short-range
four fermion interaction is irrelevant for $(3+1)d$ Dirac or
chiral fermions. Thus we assume that the interaction on the mirror
sector is stronger than the GUT, thus the interaction only gaps
out the mirror sector. Alternatively, we can take a uniform
interaction in the entire system, but make the kinetic energy
stronger on the GUT but weaker on the mirror sector.

The second criterion mentioned above implies that the $4d$ bulk TI
must be trivialized by interaction. This effect of interaction on
TI was studied immensely in condensed matter community in the last
few
years~\cite{fidkowski1,fidkowski2,qiz8,yaoz8,zhangz8,levinguz8,chenhe3B,senthilhe3,xu16,youinversion},
and now it is understood that in one, two, and three spatial
dimensions, there are examples of topological insulators which are
nontrivial in the noninteracting limit, but can be trivialized by
certain {\it well-designed} interaction, namely their boundaries
can be driven into the SCSG phase by interaction. Thus to obtain
the desired lattice regularization for GUT, we need to demonstrate
the following two results:

First of all, there is a $4d$ TI which in the noninteracting limit
has massless chiral fermions on its $3d$ boundary, and the
symmetry of the TI is precisely the same as the gauge symmetry of
the GUT, so we can couple the system to the correct gauge fields;

Second, and most importantly, under interaction the $4d$ TI must
become a trivial phase, thus its boundary can be driven into the
SCSG phase.

There are two equivalent ways to prove a TI is trivialized by
interaction: {\it (1)} one can directly show that the boundary of
the TI is driven into the SCSG phase by certain interaction; {\it
(2)} alternatively, we can also prove that the
topological-to-trivial quantum phase transition in the bulk is
``erased" by interaction, namely under interaction the ``trivial
insulator" and TI in the noninteracting limit can be connected
adiabatically to each other without closing the bulk
gap~\footnote{It is believed (although not proved) that these two
approaches are equivalent, namely if two $d-$dimensional states
can be adiabatically connected by tuning a parameter (for instance
the Dirac mass $m$) without closing the bulk gap, then it implies
that the $(d-1)-$dimensional interface between these two states
can be gapped and nondegenerate. To visualize this statement, one
can just make a smooth and wide interface, over which the tuning
parameter changes smoothly in space from one state to another,
then the gap never closes at this smooth interface. }. In section
{\bf 2}, we will apply the {\it first} approach to a toy model,
which is similar to the GUT in the sense that its $3d$ boundary
has 16 gapless chiral fermions; In section {\bf 3}, we will use
the {\it second} approach to show that the Pati-Salam GUT
\cite{patisalam} emerges as the boundary of a $4d$ TI, and the
mirror sector is decoupled in the IR because it can be driven into
the SCSG phase by interaction.

The first lesson we learned from the studies of interacting TI is
that, the SCSG phase does {\it not} exist for arbitrary flavors of
fermions. It is now well-understood that in $0d$ and $1d$, SCSG
phase only exists for $8n$ flavors of Majorana fermions with
integer $n$; in $2d$, SCSG phase only exists for $16n$ flavors of
Majorana fermions (a review of these previous results is given in
appendix A). The interaction that realizes the SCSG phase must be
a flavor mixing interaction term, whose explicit form was given in
$0d$ and $1d$~\cite{fidkowski1,fidkowski2,yoni}. Thus one has to
carefully select the short range interaction terms to realize the
SCSG phase.

We note here that the all-important SCSG phase of the mirror
sector was also sought for in the
past\cite{latticefermions,domainNarayanan,domainGolterman1,domainGolterman2,lattice345}.
This phase was first proposed in high energy physics community in
Ref.~\cite{preskill1986} and it was called the Eichten-Preskill
mechnism. But the existence of the SCSG phase was never firmly
established. Recently new proposal of constructing SCSG phase
based on classification of symmetry protected topological states
(a generalization of topological insulator) was made in
Ref.~\cite{wen2013A,wen2013B}, which is similar to the logic we
presented in this section. Besides, SCSG phases for anomaly-free
$(1+1)d$ systems were also discussed in Ref.~\onlinecite{juven}.
In Ref.~\cite{wen2013A,wen2013B}, the general diagnosis for
classification of fermionic SPT states was based on the
computation of super-cocycles of symmetry group. In our current
work, we will use a very different way of understanding
classification of interacting TIs, which is more intuitive and
more convenient to analyze compared with super-cocycle
calculation, especially for the Lie groups involved in GUT.
Meanwhile, our method not only demonstrates the existence of the
SCSG phase, but also gives us guidance for constructing the
specific interaction that realizes the SCSG phase.

\section{2. A toy model}

Let us first start with a toy model, whose bulk theory is a $4d$
TI with a $\U1$ and $\dsZ_2$ symmetry, and we can use the same
bulk band structure introduced for the $4d$ quantum Hall state in
Ref.\,\cite{qi2008,wen2013B}: \beqn H_\text{TI} &=& \sum_{a = 1}^2
\sum_{\vec{k}} \psi^\dagger_{\vec{k},a} \Big( \sum_{i = 1}^4
\Gamma^i \sin(k_i) \Big) \psi_{\vec{k},a} \cr\cr &+&
\psi^\dagger_{\vec{k},a} \Gamma^5 \Big( \sum_{i = 1}^4\cos(k_i) -
4 + m\Big) \psi_{\vec{k},a} , \label{4dbulk}\eeqn where
$\Gamma^{1,2,3} = \sigma^3\otimes\sigma^{1,2,3} $, $\Gamma^{4} =
\sigma^1\otimes \sigma^0$, $\Gamma^{5} = \sigma^2\otimes\sigma^0 $
(with $\sigma^{1,2,3}$ being the Pauli matrices and $\sigma^0$
being the $2\times2$ identity matrix). $m > 0$ and $m < 0$
correspond to the topological and trivial insulators respectively.
Close to the critical point $m = 0$, when expanded around $\vec{k}
= 0$, \eqnref{4dbulk} becomes the standard $4d$ Dirac fermion
Hamiltonian: $H_\text{TI} = \sum_{a=1,2} \sum_{\vec{k}}
\psi^\dagger_{\vec{k},a} ( \vec{k} \cdot \vec{\Gamma} + m \Gamma^5
) \psi_{\vec{k},a} $.

The $3d$ boundary of this theory (which is the domain wall of mass
$m$: Fig.~\ref{fig: 4dTI}) has precisely two flavors of chiral
fermions (domain wall fermions): \beqn H_{3d} = \int \dd^3\vect{x}
\ \sum_{a = 1}^2 \psi^\dagger_a (\ii\vect{\sigma} \cdot
\vect{\partial}) \psi_a, \label{3db} \eeqn with  $\sigma^{x,y,z}$
being the Pauli matrices in the spin space. The $\U1$ and $\dsZ_2$
symmetries act on the boundary chiral fermions as \beqn \U1 :
\psi_a \rightarrow [e^{\ii \tau^y \theta}]_{ab} \psi_b,
\quad\dsZ_2 : \psi_a \rightarrow (\tau^y)_{ab} \psi_b, \label{sym}
\eeqn where $\tau^{x,y,z}$ denote the Pauli matrices in the flavor
space.

As long as we preserve the $\U1\times \dsZ_2$ symmetry, the $3d$
boundary can never be gapped without interaction for {\it
arbitrary} copies of this system, because the only fermion
bilinear mass terms that can gap out the boundary are the Cooper
pair operators: $\psi_a^\intercal \ii \sigma^y \psi_b + H.c.$
which inevitably break at least one of the symmetries. Thus this
$4d$ TI has a $\mathbb{Z}$ classification with the $\U1\times
\dsZ_2$ symmetry, at the noninteracting level (see appendix D for
a proof of the classification).

In the following we will argue that short range interactions can
reduce the classification of this $4d$ TI to $\mathbb{Z}_{8}$:
local four-fermion interactions can gap out 8 copies of
\eqnref{3db} and drive it into a SCSG phase. Notice that here the
$\U1$ symmetry is not anomalous (it is analogous to the $B - L$
$\U1$ symmetry of the Standard Model), thus a SCSG phase in this
case does not violate the anomaly matching condition.

Directly studying strong four-fermion interactions is difficult,
so we will follow the same logic as in
Ref.\,\cite{senthilhe3,TI_fidkowski2,TI_qi,TI_senthil,TI_max,maxhe3}:
we will first manually break a subgroup of the $\U1\times \dsZ_2$
symmetry by condensing an order parameter that transforms
nontrivially under these symmetries. Then we will
condense/proliferate the defects of the condensate to restore the
broken symmetry. After condensing the defects, the order parameter
becomes disordered and can be safely integrated out. This
generates an effective interaction at low energy.

Let us first spontaneously break the $\U1$ symmetry by condensing
an O(2) ``superfluid" order parameter with unit length
$\vect{n}=(n_1,n_2)\in\dsR$ at the $3d$ boundary, which couples to
the fermions as: \beqn H_{\text{O(2)}} = \vect{n}\cdot
(\Re[\psi^\intercal \tau^x \sigma^y \psi], \,\Re[\psi^\intercal
\tau^z \sigma^y \psi]). \label{o2} \eeqn This superfluid order
parameter gaps out the chiral fermions and breaks the $\U1$
symmetry, but preserves the $\dsZ_2$ symmetry in \eqnref{sym}. The
broken $\U1$ symmetry can be restored by condensing the vortex
lines of the O(2) order parameter $\vect{n}$ in \eqnref{o2}.

The dynamics of vortex lines can be systematically described in
the dual formalism. In an ordinary $3d$ superfluid phase with
spontaneous $\U1$ symmetry breaking, the $\U1$ Goldstone mode is
dual to a rank-2 antisymmetric tensor field $B_{\mu\nu}$ defined
as \beqn J_\mu = \epsilon_{\mu\nu\rho\tau} \partial_\nu
B_{\rho\tau}. \eeqn $J_{\mu}$ is the superfluid current.
$B_{\mu\nu}$ is coupled to the vortex loops, which now can be
described by a vector gauge field $a_\mu$, and the fact that the
vortex loop can never end corresponds to the Gauss law of the
gauge field $\vect{\nabla} \cdot \vect{e} = 0$. The dynamics of
vortex loops can be described by the following schematic dual
action in the $4d$ Euclidean space lattice (which corresponds to
the $(3+1)d$ boundary space-time): \beqn \mathcal{S} &=&
\sum_{\vec{x}} - t\cos(\nabla_\mu a_\nu - \nabla_\nu a_\mu -
B_{\mu\nu}) \cr\cr &-& \frac{1}{e^2} (\epsilon_{\mu\nu\rho}
\nabla_\mu B_{\nu\rho})^2 + \cdots \eeqn The details of this
standard duality formalism can be found in Ref.\,\cite{savit}.
Depending on $t / e^2$, this model has two different phases: with
$t / e^2\ll 1$ the vortex loops are ``small", and the system has
one gapless gauge boson $B_{\mu\nu}$, which is the dual of the
Goldstone mode in the superfluid phase; while with $t / e^2 \gg
1$, the vortex loops condense, and $B_{\mu\nu}$ and $a_\mu$ will
both be gapped due to their coupling, this corresponds to the
quantum disordered phase of superfluid.

The above dual action only applies when the vortex loop is
``trivial", namely there is no extra low energy degree of freedom
besides the vortex loops and superfluid Goldstone mode. This
requires the fermion ground state with a vortex loop background be
gapped and nondegenerate. For example, if the fermion ground state
within a vortex loop is two fold degenerate, then the vortex loop
will carry an extra flavor index $i = 1,2$. In this case after the
vortex loops condense, the system is {\it not} fully gapped,
instead, it would enter a phase with a gapless photon
excitation~\cite{senthilloop}; or in other words, after the
superfluid order parameter is disordered, this system becomes a
``$\U1$ spin liquid" phase. This is because $a_{1,\mu}$ and
$a_{2,\mu}$ are both coupled to $B_{\mu\nu}$, thus when both
$a_{1,\mu}$ and $a_{2,\mu}$ proliferate, $a_{+, \mu} = a_{1,\mu} +
a_{2,\mu}$ will be rendered gapped by $B_{\mu\nu}$, while
$a_{-,\mu} = a_{1,\mu} - a_{2,\mu}$ remains gapless since it is
not coupled to any dual Goldstone mode. More details about quantum
phases after proliferation of degenerate vortex loops can be found
in Ref.\,\cite{senthilloop}.

Thus, \textit{the desired SCSG phase is only possible when the
defects in the condensate have a trivial spectrum.} we have to be
careful with the core of the vortex line, since it is the
singularity of the O(2) order parameter, and the fermions may
become gapless along the vortex line. Now we have reduced our
original $3d$ problem to a $1d$ problem inside a vortex line,
which we can analyze much more reliably. In our current case, the
vortex line of this O(2) order parameter traps $1d$ nonchiral
Majorana fermion modes that are localized at the vortex
line.\cite{teo} Upon solving the Dirac equation in the vortex
background, we find that these modes are described by the
Hamiltonian: \beqn H_{1d} = \frac{1}{2}\int \dd x\,( \chi_L \ii
\partial_x \chi_L - \chi_R \ii \partial_x \chi_R ). \label{1db}
\eeqn and their transformation properties under the residual
$\dsZ_2$ symmetries are: \beqn \dsZ_2 : \chi_L \rightarrow \chi_L,
\ \ \chi_R \rightarrow - \chi_R. \label{1dsym}\eeqn With this
symmetry, it is straightforward to verify that for arbitrary
numbers of the $1d$ system \eqnref{1db}, any fermion bilinear mass
term is forbidden. For example, $\bar{\chi} \chi = 2 i \chi_L
\chi_R $ is forbidden by the $\dsZ_2$ symmetry \eqnref{1dsym},
thus without interaction, this $1d$ system cannot be gapped
without degeneracy, for arbitrary copies of this system
\eqnref{1db}, then this implies that without turning on certain
interaction at the vortex core, a SCSG phase can {\it not} be
obtained by condensing the vortex loops.

However, Ref.\,\cite{fidkowski1,fidkowski2,qiz8} showed that
although all the fermion bilinear mass terms are forbidden in
\eqnref{1db}, when there are $8n$ copies of \eqnref{1db}, a
particular four fermion interaction term which preserves the
$\dsZ_2$ symmetry still gaps out the $1d$ fermions with $\langle
\bar{\chi}_a \chi_b \rangle = 0$ for arbitrary flavor index $a,b$.
The specific form of this interaction was given in
Ref.\,\cite{fidkowski1,fidkowski2,yoni} and reviewed in appendix
B, and it can also be concisely written as
\begin{equation}
H_\text{int}=-\frac{g}{2}\int \dd x\,\sum_{a=1}^7 (\chi_L^\intercal
\gamma^a\chi_R+\text{H.c.})^2 , \label{eq: int 1d SO(7)}
\end{equation}
where the Majorana field $\chi_{L,R}$ has been extended to
eight-component. The coupling matrices $\gamma^a$ are the Gamma
matrices of the $\SO7$ group in its 8-dimensional spinor
representation, which, under a specific choice of basis, may be
written as
$\vect{\gamma}=(\sigma^{002},\sigma^{323},\sigma^{021},\sigma^{203},\sigma^{231},\sigma^{123},\sigma^{211})$
(hereinafter $\sigma^{ijk\cdots} \equiv
\sigma^i\otimes\sigma^j\otimes\sigma^k\cdots$ denotes the tensor
product of Pauli matrices). As proven in
Ref.\,\cite{fidkowski1,fidkowski2,yoni}, such interaction can
drive eight copies of the 1d system \eqnref{1db} into a SCSG phase
at strong coupling. Then the O(2) vortex loops can condense to
restore the $\U1$ symmetry and gap out the chiral fermions on the
$3d$ boundary.

To explicitly implement our picture of condensing vortex loops, we
need to control the dynamics of the vortex loops. In order to do
this, we propose to add the following interacting Hamiltonian on
the $4d$ lattice model: \beqn H_{\mathrm{total}} = H_\text{int-4d}
+ H_{\text{O(2)}} + H[\vect{n}]. \eeqn $H_{\mathrm{int}}$ is a
four-fermion interaction term which generates the \eqnref{eq: int
1d SO(7)} in every vortex loop, which will gap out the vortex loop
without degeneracy. Its explicit form in the $4d$ bulk and at the
$3d$ boundary reads (see appendix B for derivation): \beqn
H_\text{int-4d} &=& - \frac{g}{2}\int \dd^4 \vect{x} \sum_{a=1}^7
\Re[\psi^\intercal \tau^y \Gamma^2 \gamma^a \psi]^2, \cr\cr
H_\text{int-3d} &=& -\frac{g}{2}\int \dd^3\vect{x} \sum_{a=1}^7
\Re[\psi^\intercal \tau^y \sigma^y \gamma^a \psi]^2.\label{eq: int
3d SO(7)} \eeqn $H_{\text{O(2)}}$ is the coupling between the O(2)
vector $\vect{n}$ to the fermions on the lattice model, which
generates coupling \eqnref{o2} on the $3d$ boundary. $H[\vect{n}]$
controls the dynamics of the O(2) vector $\vect{n}$, including the
dynamics of the vortex loops. We parametrize $\vect{n}$ as
$\vect{n} = (\cos(\hat{\phi}), \sin(\hat{\phi}))$, where
$\hat{\phi} \in [0, 2\pi)$, and label the canonical momentum of
$\hat{\phi}$ as $\hat{N}$, with $\hat{N} \in \text{Integers}$. We
propose the following Hamiltonian $H[\vect{n}]$: \beqn H[\vect{n}]
&=& \sum_{\vect{x}, \mu \neq \nu} - J \cos\big( \nabla_\mu
\hat{\phi} \big)  + V[\hat{N}(\vect{x})] \cr \cr &+& K \cos\big(
\nabla_{\mu} \nabla_{\nu} \hat{\phi} \big), \eeqn where the sum is
taken over all spatial positions $\vect{x}$ and directions $\mu$,
$\nu$. The lattice derivatives are defined as $\nabla_\mu
\hat{\phi}(\vect{x}) = \hat{\phi}(\vect{x} + \mu) -
\hat{\phi}(\vect{x}) $, $\nabla_\mu\nabla_\nu \hat{\phi}(\vect{x})
= \hat{\phi}(\vect{x} + \mu + \nu) - \hat{\phi}(\vect{x} + \nu) -
\hat{\phi}(\vect{x} + \mu) + \hat{\phi}(\vect{x}) $. $V[\hat{N}]$
is a local short range repulsive interaction of $\hat{N}$, whose
explicit form has many choices, but the simplest possibility is
$V[\hat{N}(\vect{x})] = v \big(\hat{N}(\vect{x})\big)^2 $. When
$J$ dominates all the other terms, $\vect{n}$ is ordered, the O(2)
symmetry is spontaneously broken, and the fermions acquire an
ordinary fermion gap. If we start with a weak superfluid phase (a
superfluid phase with a small stiffness), the $K$ term will
compete with the superfluid order by lowering the core energy of
vortices, and we expect it to drive the system into a vortex
condensate, with an appropriate choice of $V[\hat{N}]$. An
analogue of $H[\vect{n}]$ in $2d$ was studied by quantum Monte
Carlo in Ref.~\cite{sandvik1,sandvik2}. It was shown in a
spin-1/2 quantum XY model that when the ring exchange term $K$
dominates $J$, it indeed drives a order-disorder quantum phase
transition. Thus the $K$ term can be effectively viewed as $ \sim
- K \rho_v^2$, where $\rho_v$ is the local density of vortices.

Our toy model demonstrated that eight copies of the $4d$ TI
\eqnref{4dbulk} can be trivialized under a local fermion
interaction with $\SO7\times\SO2$ symmetry. Since the mirror
sector is driven into the SCSG phase, we can obtain 16 chiral
fermions on the other $3d$ boundary with lattice regularization.
In fact, the symmetry group can be further enlarged to
$\SO7\times\SO3$. In that case, we introduce an O(3) order
parameter with unit length $\vect{n}=(n_1,n_2,n_3)\in\dsR$ which
couples to the boundary fermions as
$H_{\text{O(3)}}=\vect{n}\cdot(\Re[\psi^\intercal \tau^x \sigma^y
\psi], \,\Re[\psi^\intercal \tau^z \sigma^y
\psi],\,\Im[\psi^\intercal \sigma^y \psi])$. Following the similar
defect condensation argument, we can first gap out the chiral
fermions on the $3d$ boundary by ordering the O(3) order parameter
at the price of breaking the $\SO3$ symmetry, and then we attempt
to restore the symmetry by condensing the monopole defects of
$\vect{n}$. Each monopole will trap eight Majorana zero modes
$\chi$ (the calculation is identical to that in
Ref.~\cite{jackiw1976}), which can not be gapped out by any
fermion bilinear terms because they are all forbidden by the
$\SO7$ symmetry. Now the same $3d$ interaction in \eqnref{eq: int
3d SO(7)} will induce the following $0d$ interaction among the
eight Majorana zero modes at the monopole core:
\begin{equation}\label{eq: int 0d SO(7)}
H_\text{int}=-\frac{g}{2} \sum_{a=1}^7
(\chi^\intercal\gamma^a\chi)^2,
\end{equation}
with the same set of $\gamma^a$ matrices defined below \eqnref{eq:
int 1d SO(7)}. As shown in Ref.\,\cite{fidkowski1,fidkowski2,yoni}
and reviewed in appendix B, this $0d$ interaction can gap out the
Majorana zero modes and stabilize a unique $\SO7$ singlet ground
state in the monopole core. It can also be verified that the
monopole defect is a boson, so it can condense to restore the
$\SO3$ symmetry. Thus the chiral fermions in the mirror sector can
also be driven into the SCSG phase with the larger symmetry
$\SO7\times\SO3$ as well.

One can see that the symmetry group and the design of the
interaction may vary from case to case, but the common features
that we wish to emphasize are: \emph{(1)} the interaction terms we
turn on explicitly breaks the anomalous $\U1$ symmetry of the
boundary chiral fermions, thus a SCSG phase is possible;
\emph{(2)} the counting of 16 chiral fermions is crucial, if the
fermion flavor number is insufficient, the SCSG phase will not be
realized and the mirror sector can not be decoupled by
interaction.

The key of the analysis in this section is to show that a properly
designed $4d$ bulk interaction can induce the {\it correct} $1d$
($0d$) four fermion interaction \eqnref{eq: int 1d SO(7)}
(\eqnref{eq: int 0d SO(7)}) inside the vortex loop (monopole
core), which is known to be capable of driving the vortex loop
(monopole core) into a SCSG phase~\cite{fidkowski1,fidkowski2}. In
the next section, we will also use the dimensional reduction
argument, and we show that the Standard Model can be successfully
regularized as part of the Pati-Salam GUT on the boundary of a
$4d$ TI, and the mirror sector can be driven into the SCSG phase
and hence decoupled in the IR.

\section{3. Pati-Salam GUT}

Motivated by the Pati-Salam GUT whose gauge group is
$\SU4\times\SU2_1\times\SU2_2$, we may directly start from a $4d$
TI with $\SU4\times\SU2_1\times\SU2_2$ as its symmetry group. The
lattice model of the $4d$ TI is of the same form as
\eqnref{4dbulk}, expect that now $\psi_{\vec{k},a}$ (for each
$a=1,2$ respectively) is extended to an eight-flavor Dirac fermion
field. The Hamiltonian respects the $\SU4\times\SU2_1\times\SU2_2$
symmetry in the way that $\psi_{\vec{k},1}$ and $\psi_{\vec{k},2}$
form the representations $(\mathbf{4},\mathbf{2},\mathbf{1})$ and
$(\bar{\mathbf{4}},\mathbf{1},\mathbf{2})$ respectively. Its 3d
boundary theory still takes the same form as \eqnref{3db}, but the
boundary fermions $\psi_a$ ($a=1,2$) now transform under
$\SU4\times\SU2_1\times\SU2_2$ like
$(\mathbf{4},\mathbf{2},\mathbf{1})$ and
$(\bar{\mathbf{4}},\mathbf{1},\mathbf{2})$ representations, which
can be written out explicitly as
\beqn{\renewcommand\arraystretch{1.6}\begin{array}{rll} \SU4: &
\psi_1\to e^{\ii\vect{\theta}\cdot\vect{\rho}}\psi_1, & \psi_2\to
e^{-\ii\vect{\theta}\cdot\vect{\rho}^\ast}\psi_2;\\
\SU2_1: & \psi_1\to
e^{\ii\vect{\theta}_1 \cdot\vect{\mu}}\psi_1,& \psi_2\to\psi_2;\\
\SU2_2: & \psi_1\to\psi_1,& \psi_2\to e^{\ii\vect{\theta}_2
\cdot\vect{\mu}}\psi_2.
\end{array}}\eeqn
$\vect{\rho}$ and $\vect{\mu}$ denote the generators of $\SU4$ and
$\SU2$ groups respectively.

As long as the $\SU4\times\SU2_1\times\SU2_2$ symmetry is
preserved, the $3d$ boundary must remain gapless at the
free-fermion level. Because all the fermion bilinear mass terms at
the $3d$ boundary take the form of the spin-singlet Cooper
pairing: $\psi_a^\intercal \ii\sigma^y M \psi_b + H.c.$ (where
$a,b=1,2$ and $M$ is an arbitrary matrix in the color-flavor
space), but such terms are forbidden by the $\SU4$ symmetry if
$a=b$, and are forbidden by the $\SU2_1\times\SU2_2$ symmetry
if $a\neq b$, therefore no fermion bilinear mass term can be added
without breaking the $\SU4\times\SU2_1\times\SU2_2$
symmetry. Thus the $4d$ insulating phases with
$\SU4\times\SU2_1\times\SU2_2$ symmetry is $\dsZ$
classified (see appendix D for a proof of the
classification).

Another way of making the same statement is to say that, at the
free-fermion level, the $4d$ bulk TI can not be smoothly tuned
(while preserving the symmetry) into a trivial insulator without
going through a gap-closing phase transition. Tuning the TI to
trivial corresponds to driving the mass $m$ of a bulk $4d$ Dirac
fermion from positive to negative, and close to the quantum
critical point $m = 0$ and expanded at $\vec{k} = 0$, the bulk
theory reads: \beqn H_\text{TI} = \int \dd^4\vect{x} \sum_{a =
1,2} \psi^\dagger_{a} (\ii \vec{\Gamma} \cdot \vec{\partial} + m
\Gamma^5 )\psi_{a}, \label{4dDirac} \eeqn where $\psi_a$ for each
$a$ is an eight-flavor Dirac fermion which also carries $\SU4$ and
$\SU2$ indices. While changing $m$, the fermion bulk gap will
close at $m=0$. Without interaction, the gap-closing transition
can not be circumvented, because there is no other
symmetry-allowed mass terms to be added that can gap out the point
$m=0$. For example, the Majorana mass terms $\psi_a^\intercal
\ii\Gamma^2 M \psi_b+H.c.$ could gap out the bulk criticality at
$m=0$, however, as we have shown before, such terms are all
forbidden by the $\SU4\times\SU2_1\times\SU2_2$ symmetry. So
without interaction the $4d$ $\SU4\times\SU2_1\times\SU2_2$ TI and
the trivial insulator are in different phases, separated by a
phase transition that can not be avoid at the non-interacting
level without breaking the symmetry.

However, as seen before (and also recently studied in
literatures\cite{fidkowski1,fidkowski2,qiz8,zhangz8,levinguz8,yaoz8,chenhe3B,senthilhe3,xu16,youinversion}),
the classification of topological insulators can be reduced by
interaction. Here, as we will show in the following, the
classification of the $4d$ $\SU4\times\SU2_1\times\SU2_2$ TI is
reduced from $\dsZ$ to trivial, meaning that under interaction the
$4d$ TI and the trivial insulator are actually in the same phase,
and the bulk phase transition between them can be avoid by
strong-enough and properly-designed interactions, as shown in the
phase diagram \figref{fig: levels}(a). In other words, the gapless
bulk fermion at the $m=0$ critical point can be gapped out by
interaction. %\cite{doubleQSH, shailesh}

\begin{figure}[htb]
\begin{center}
\includegraphics[width=220pt]{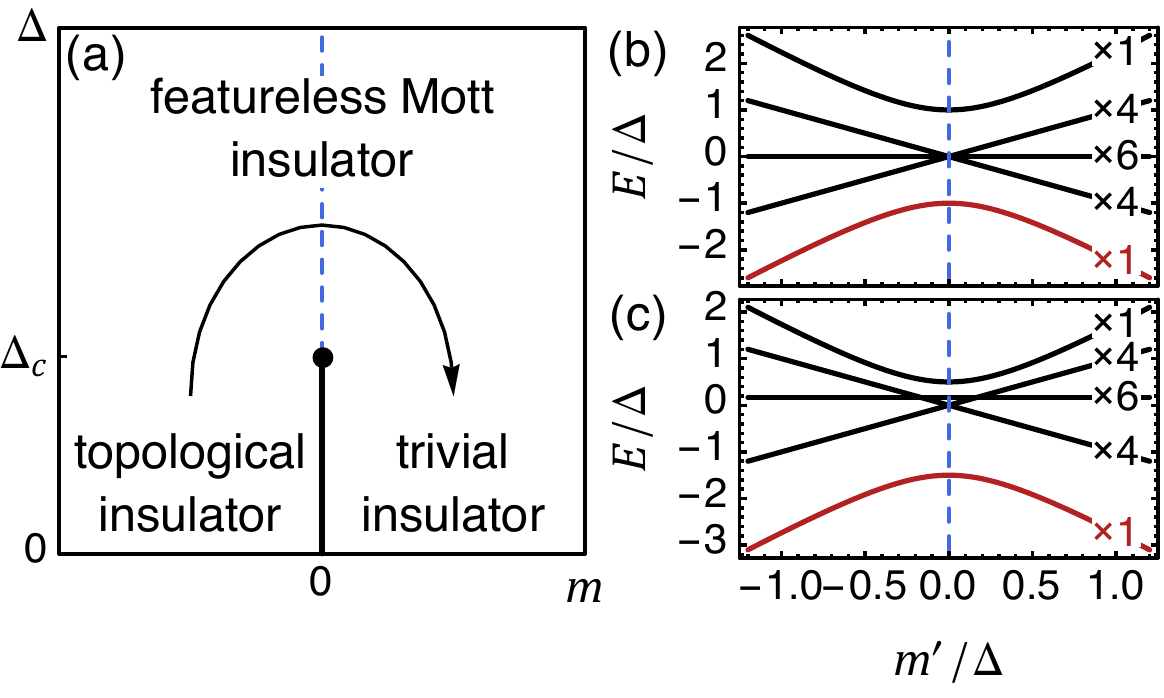}
\caption{(a) Schematic phase diagram of the $4d$ TI with
$\SU4\times\SU2_1\times\SU2_2$ symmetry under interaction.  There
exist a critical interaction strength $\Delta_c$, above which the
topological-to-trivial transition can be circumvented. (b,c) The
O(4) monopole core levels along a path connecting the $4d$ TI to
the trivial insulator, parameterized by the reduced mass $m'$. The
effective Hamiltonian in the monopole core reads
$H=H_\text{free}+H_\text{int}$, where $H_\text{int}$ is taken from
(b) \eqnref{eq: int core} or (c) \eqnref{eq: int fact}. The
16-dimensional Hilbert space split according to $\SU4$
representations as
$\mathbf{16}=\mathbf{1}+\mathbf{1}'+\mathbf{4}+\bar{\mathbf{4}}+\mathbf{6}$
with the unique ground state $\ket{\mathbf{1}}+\ket{\mathbf{1}'}$
(marked out in red). The dashed line marks out the $m'=0$ critical
point, where degeneracy is avoided by interaction.} \label{fig:
levels}
\end{center}
\end{figure}

To show this, we will again implement the argument of defect
proliferation/condensation, i.e. one may choose to break part of
the symmetry by condensing certain fermion-bilinear order
parameter, and then restore the symmetry by condensing topological
defects of that order parameter field. Due to the fact
$\SU2_1\times\SU2_2\simeq\SO4$, one can introduce the
symmetry-breaking O(4) vector order parameter field
$\vec{n}=(n_0,\vect{n})=(n_0,n_1,n_2,n_3)\in\dsR$, which couples
to the bulk fermions as
\begin{equation}\label{eq: O4}
H_{\text{O(4)}}=\int d^4\vect{x}\,n_0 \psi_1^\intercal \ii
\Gamma^2 \psi_2+\vect{n}\cdot \psi_1^\intercal \Gamma^2 \vect{\mu}
\psi_2 + \text{H.c.}.
\end{equation}
The $\SU2_1$ and the $\SU2_2$ rotations act respectively as the
left and the right isoclinic rotations on the O(4) vector
$\vec{n}$. We first condense $\vec{n}$ to gap out the bulk
fermions for all range of $m$ (including $m=0$) at the price of
breaking the $\SU2_1\times\SU2_2$ symmetry. In the $4d$ bulk, a
O(4) vector order parameter has the hedgehog monopole topological
defects, due to the fact $\pi_3[S^3] = \mathbb{Z}$. The broken
symmetry is expected to be restored by condensing the O(4) vector
monopole defects in the $4d$ bulk. Since our goal is to show that
the critical point $m = 0$ can be gapped out by interaction, we
only need to demonstrate that under interaction the fermion
spectrum inside the monopole will always be gapped and
nondegenerate in the entire phase diagram.

By directly solving the Schr\"odinger equation (see appendix C for
details), one can show that the monopole will trap four complex
fermion localized modes $f_i$ ($i=1,2,3,4$) forming a fundamental
representation of the $\SU4$ symmetry. Since the $\SU4$ symmetry
is not broken by the O(4) vector $\vec{n}$, the effective
Hamiltonian of $f_i$ must be $\SU4$ invariant. Without
interaction, the only fermion bilinear Hamiltonian reads \beqn
H_{\mathrm{free}} \sim m^\prime \sum_{i = 1}^4 (f^\dagger_i f_i -
1/2), \label{monopolefree}\eeqn here the coefficient $m^\prime$ is
proportional to the mass $m$ of the bulk Dirac fermion. Thus by
tuning $m$ from negative to positive, the monopole core will close
its spectrum gap at $m \sim m^\prime = 0$, and the monopole will
be 16-fold degenerate at $m^\prime = 0$. Thus without four-fermion
interaction inside the monopole, condensing the monopole will
still lead to a bulk quantum phase transition at $m =
0$~\footnote{If we project the $4d$ Dirac fermion \eqnref{4dDirac}
to the monopole core, then $m = 0$ in the $4d$ bulk precisely
coincides with $m^\prime = 0$ in the monopole core; the parameter
$m$ in the lattice model will not be exactly proportional to
$m^\prime$ in \eqnref{monopolefree}, but since $H_{\mathrm{free}}$
is the only noninteracting term inside a monopole core, tuning $m$
from negative to positive in the lattice model will definitely
cross the point $m^\prime = 0$.}.

However, the monopole core spectrum can be completely changed by
the following $\SU4$ invariant local four-fermion interaction
\begin{equation}\label{eq: int core}
H_\text{int}=-\Delta(f_1f_2f_3f_4+\text{H.c.}),
\end{equation}
At $m^\prime = 0$, $H_{\mathrm{int}}$ will lift the degeneracy
among these fermion zero modes, and single out the unique ground
state $(\ket{0000}+\ket{1111})/\sqrt{2}$. $\ket{0}$ and $\ket{1}$
stand for the fermion occupation number eigenstates of the zero
mode. A slightly different interaction (see appendix B for
derivation) will play qualitatively the same role as \eqnref{eq:
int core}:
\begin{equation}\label{eq: int fact}
\begin{split}
H_\text{int}=&-\frac{g}{2}(f^\intercal \vect{\lambda} f + \text{H.c.})^2\\
=&-24 g(f_1f_2f_3f_4+\text{H.c.})-8g\sum_{i<j}\rho_i\rho_j,\\
\vect{\lambda}=&(\sigma^{12},\sigma^{20},\sigma^{32},-\ii\sigma^{21},-\ii\sigma^{02},-\ii\sigma^{23}),
\end{split}
\end{equation}
where $\lambda^a$ ($a=1,\cdots,6$) are six $4\times4$ matrices
acting in the $\SU4$ color sector (forming the representation
$\mathbf{6}$ of $\SU4$), and $\rho_i=f_i^\dagger f_i-\frac{1}{2}$
($i=1,2,3,4$) denote the fermion density operators. The first term
is the interaction in \eqnref{eq: int core} by identifying
$\Delta=24 g$, and the second term is a density-density
interaction which does not qualitatively change the spectrum of
monopole, as seen by comparing \figref{fig: levels}(b,c).

%\cite{latticefermions}
%The interaction drives the system into a charge-4 superconductor
%with a many-body gap $\Delta$.

With the protection of the gap $\Delta$, the ground state of the
O(4) monopole evolves smoothly in all range of $m$ without any
level crossing with the excited states, as shown in \figref{fig:
levels}(b,c). The interaction not only renders the monopole to a
nondegenerate $\SU4$ singlet, it also makes the monopole a boson,
this is because deep in the trivial phase of \eqnref{4dbulk} and
\eqnref{4dDirac}, $i.e.$ when $m$ is negative and it is the
dominant energy scale of the system, the ground state of a
monopole in the bulk must be a featureless boson. And we have
proved that with interaction $H_{\text{int}}$ the ground state of
the monopole never has any level-crossing with excited states,
thus the ground state of the monopole must remain as a boson for
the entire range of $m$. Thus the monopole can be safely condensed
to restore the broken $\SU2_1\times\SU2_2$ symmetry without
causing ground state degeneracy or breaking other symmetries.
After the monopole condensation, we end up with a symmetric gapped
phase in the bulk for the entire range of $m$, meaning that the
bulk phase transition between the $4d$
$\SU4\times\SU2_1\times\SU2_2$ TI and the trivial insulator can be
removed by the interaction with sufficient strength, see
\figref{fig: levels}(a). Note that the $3d$ boundary of the $4d$
TI is simply the spatial interface between the $4d$ TI and the
trivial insulator (vacuum). Since the $4d$ TI can now smoothly
evolve into the trivial insulator without gap-closing phase
transition, the $3d$ interface between these two states (which can
be viewed as an evolution in space) must also be driven into a
SCSG phase by the same kind of interaction.

Again, to explicitly implement our picture of ``condensing
topological defects", we need to control the dynamics of the
topological defects. In order to do this, we propose to add the
following interacting Hamiltonian on the $4d$ lattice model: \beqn
H_{\mathrm{total}} = H_\text{int-4d} + H_{\text{O(4)}} +
H[\vec{n}]. \eeqn $H_\text{int-4d}$ is a $4d$ bulk interaction
that will induce the correct four-fermion term at the monopole
core, which gaps out the monopole for all range of $m$ in the
phase diagram (see appendix B for derivation):
\begin{equation} H_\text{int}=-\frac{g}{2}\int
\dd^4\vect{x}\; (\psi_1^\intercal \Gamma^2\mu^2 \vect{\lambda}
\psi_1 + \psi_2^\dagger \Gamma^2\mu^2 \vect{\lambda}
\psi_2^{\dagger\intercal} + \text{H.c.})^2. \label{eq: int 4d
SO(6)}
\end{equation}
This interaction is manifestly $\SU4\times\SU2_1\times\SU2_2$
invariant. $H_{\text{O(4)}}$ is given by \eqnref{eq: O4}.
$H[\vec{n}]$ is the Hamiltonian for the O(4) unit vector order
parameter $\vec{n}$ that should control the dynamics of $\vec{n}$
and its topological defects: \beqn H[\vec{n}] = \sum_{\vect{x},
\mu} - J \left( \nabla_\mu \vec{n} \right)^2 + V[L^{ab}(\vect{x})]
- K \rho_m(\vect{x})^2 . \eeqn $V$ is a local interaction between
SO(4) angular momentum operator $L^{ab}(x)$ (which is conjugate to
operator $\vec{n}(\vect{x})$), its simplest form could be $\sum_{a
< b}v \big( L^{ab}(\vect{x}) \big)^2 $. When $J$ dominates all the
other terms, vector $\vec{n}$ would be ordered, and the fermions
will acquire an ordinary mass gap. $\rho_m(\vect{x})$ is the local
monopole density of the SO(4) vector $\vec{n}$. To define a
monopole on a lattice, one can just follow the strategy of
Ref.\,\cite{lesikashvin}, which defined $\SO3$ monopole on a $3d$
cubic lattice and numerically studied its effects on phase
transitions. Thus we can start with a weak order of $\vec{n}$
(when $J$ and $V$ terms are comparable with each other), and
gradually increasing $K$. Then we expect that across a finite
critical point, the $K$ term will drive the system into a monopole
condensate in the $4d$ bulk. And based on our argument presented
before, the same interaction can drive the mirror sector on the
$3d$ boundary into the desired SCSG phase.

Normally condensing a conserved bosonic point particle will lead
to a gapless Goldstone mode. But in our case, inside an ordered
phase of $\vec{n}$, monopoles have long range interaction, and the
condensate of bosons with long range interaction can still have a
gapped spectrum. This is precisely the Higgs mechanism. For
example, condensing the vortices of a $(2+1)d$ superfluid will not
lead to any gapless Goldstone mode, because in the standard dual
formalism the vortex field is a complex boson which are coupled to a
dual $\U1$ gauge field.

Our analysis above suggests that if we want to regularize the
Pati-Salam GUT on a $3d$ lattice (a $4d$ lattice with a thin
fourth dimension and a decoupled mirror sector), then a
four-fermion interaction $H_{\text{int-4d}}$ is necessary. This
four-fermion interaction creates/annihilates a four-fermion $\SU4$
singlet, thus breaks the baryon number ($B$) and lepton number
conservation ($L$), but it still conserves $B-L$. For instance
this four fermion term contains the standard dimension-6 operators
that would lead to proton-decay: $qqql/\Lambda^2$. But here the UV
cut-off $\Lambda$ should be the lattice scale, which is higher
than any other scale of the system. Thus the proton decay effect
is expected to be much smaller than that predicted in the $\SU5$
GUT, which is suppressed by factor $1/\Lambda_{GUT}^2$.

\section{4. Summary}

In this work we apply the latest progress in condensed matter
physics towards understanding strongly interacting topological
insulator to the long standing problem in high energy physics: How
to regularize the SM or GUT on a lattice. In our approach, because
the bulk topological insulator is trivialized by interaction, the
mirror sector is in the SCSG phase and hence decoupled from the
GUT in the infrared limit.
%We note that similar SCSG phases,
%though not fully established, were also pursued previously in
%literature\cite{preskill1986,latticefermions,domainNarayanan,domainGolterman1,domainGolterman2,lattice345}.
Our current work heavily relies on the analysis of classification
of topological insulators under interaction,
%(which is an active new emerging field in condensed matter),
and our argument of
topological defects condensation leads to explicit construction of
an interacting lattice Hamiltonian, whose low energy physics is
described by the Pati-Salam GUT.

%\nts{We may also want to conjecture about $\SO{10}$ here.}

The authors are supported by the the David and Lucile Packard
Foundation and NSF Grant No. DMR-1151208.

\bibliography{GUT}

\clearpage
\onecolumngrid
\section{Supplemental Material}
\maketitle
\appendix
%\newcites{}{}

\section{A. Review of SCSG Phase and Interacting TI in Lower Dimensions}

The interacting fermionic topological insulator/superconductor
(TI/TSC) has recently attracted much research attention in
condensed matter physics. In the non-interacting limit, the TI/TSC
has a fully gapped and non-degenerated bulk state with gapless
fermionic boundary modes protected by symmetry. The gapless
boundary fermions are also known as the domain wall
fermions\cite{domainwall,domainKaplan,domainNarayanan,domainGolterman1}
in high energy physics. It was first pointed out by Fidkowski and
Kitaev\cite{fidkowski1,fidkowski2} that the classification of the
TI/TSC can be reduced by the fermion interaction, namely certain
non-trivial TI/TSC phases can actually be smoothly connected to
the trivial phase under interaction, and correspondingly, their
gapless boundaries can be driven into the strongly coupled
symmetric gapped (SCSG) phase by the same interaction.

\subsection{A1. SCSG Phase in $0d$: Boundary of $1d$ Systems}

The simplest example is to consider the $0d$ boundary of a $1d$
TSC, which hosts Majorana fermion zero modes (the $0d$ analog of
the domain wall fermions). The Majorana modes are denoted by the
operators $\chi_a$ ($a=1,\cdots,N$) satisfying
$\{\chi_a,\chi_b\}=2\delta_{ab}$. Let us define a time-reversal
symmetry $\dsZ_2^T$ ($\mathcal{T}^2=+1$), which acts trivially on
the Majorana modes as $\dsZ_2^T:\chi_a\to\mathcal{K}\chi_a$, where
$\mathcal{K}$ is the complex conjugation operator (flipping the
imaginary unit as $\mathcal{K}^{-1}\ii\mathcal{K}=-\ii$). Any
fermion bilinear operator $\ii\chi_a\chi_b$ will break the
time-reversal symmetry, because $\chi_a$ transforms trivially but
$\ii$ gets a minus sign. So if we require the time-reversal
symmetry, all the fermion bilinear terms will be ruled out from
the $0d$ boundary Hamiltonian, and the $0d$ boundary fermions can
not be gapped out in the free fermion limit no matter how many
modes $N$ there are. However the four-fermion interaction term
will not break the time-reversal symmetry, since no factor $\ii$
will be involved. For $N=4$, the only interaction term that one
can write down is $H=- J \chi_1\chi_2\chi_3\chi_4$. Pairing up the
Majorana fermions to regular (complex) fermions
$c_\uparrow=(\chi_1+\ii\chi_2)/2$,
$c_\downarrow=(\chi_3+\ii\chi_4)/2$, and define the fermion number
operator $n_\sigma=c_\sigma^\dagger c_\sigma$, then the
interaction Hamiltonian can be written as $H= J
(2n_\uparrow-1)(2n_\downarrow-1)$, which can be interpreted as a
Hubbard interaction leading to a two-fold degenerated ground state
(as a spin-1/2 doublet), for either sign of $J$. So if we have
$N=8$ Majorana zero mode, under the interaction, $\chi_{1,2,3,4}$
form a doublet and $\chi_{5,6,7,8}$ form another doublet, and the
two doublets can be coupled together into a singlet (such as via
the Heisenberg coupling), and the ground state degeneracy is
completely removed by the interaction, which also implies that the
expectation value of any fermion bilinear operator must vanish,
because otherwise the ground state would be degenerated. The
explicit form of the interaction is given by Fidkowski and
Kitaev\cite{fidkowski1,fidkowski2}
\begin{equation}\label{eq: int FK original}
\begin{split}
H_\text{FK} \sim &+\chi_1\chi_2\chi_3\chi_4+\chi_1\chi_2\chi_5\chi_6+\chi_1\chi_2\chi_7\chi_8+\chi_1\chi_3\chi_5\chi_7-\chi_1\chi_3\chi_6\chi_8-\chi_1\chi_4\chi_5\chi_8-\chi_1\chi_4\chi_6\chi_7\\
&-\chi_2\chi_3\chi_5\chi_8-\chi_2\chi_3\chi_6\chi_7-\chi_2\chi_4\chi_5\chi_7+\chi_2\chi_4\chi_6\chi_8+\chi_3\chi_4\chi_5\chi_6+\chi_3\chi_4\chi_7\chi_8+\chi_5\chi_6\chi_7\chi_8.
\end{split}
\end{equation}
This interaction looks rather involved and has a very high $\SO7$
symmetry, nevertheless it is not the only choice. There exist many
other interactions (to be reviewed in Appendix B) that can also
gap out eight Majorana fermions in $0d$. The point is that in this
$0d$ fermion system, only when we have eight flavors of Majorana
fermions, we can get a fully gapped spectrum and a non-degenerate
ground state (i.e. a SCSG state). If the flavor number is
insufficient ($N<8$), the Majorana zero modes can not be
completely gapped out.

The above $0d$ system is actually first realized as the boundary
of the $1d$ Majorana fermion chain\cite{kitaevchain} with the
$\mathcal{T}^2=+1$ time-reversal symmetry (in the
BDI\cite{ludwigclass1,ludwigclass2} symmetry class). The model
Hamiltonian is defined on a $1d$ lattice of the length $L$, as
$H=-\sum_{i=0}^{L-1} \ii u_i \chi_{i}\chi_{i+1} $, where $\chi_i$
($i=1,\cdots,L$) denotes the Majorana fermion operator on the site
$i$, and the bond strength $u_{i}=1+(-1)^i\delta$ alternates along
the chain, similar to the pattern of polyacetylene. If $\delta>0$,
both the bulk and the boundaries are fully gapped, and the system
is in its trivial phase; if $\delta<0$, the bulk is still fully
gapped, but each boundary will host a dangling Majorana zero mode,
and the system is in its non-trivial phase (known as a $1d$ TSC of
BDI class). The time-reversal symmetry acts as
$\dsZ_2^T:\chi_i\to\mathcal{K}(-1)^i\chi_i$, which will reduce on
the boundary to precisely the same the time-reversal symmetry that
we defined in the previous $0d$ example. Protected by the
time-reversal symmetry, the boundary Majorana zero modes can not
be gapped out at the free fermion level, and the $1d$ TSC is
therefore $\dsZ$ classified in the absence of interaction.
However, if we stack eight copies of the $1d$ TSC's together, the
boundary Majorana zero modes (now there are eight zero modes) can
be gapped out by the Fidkowski-Kitaev interaction \eqnref{eq: int
FK original} without breaking the time-reversal symmetry, and the
classification of the $1d$ TSC is reduced from $\dsZ$ to $\dsZ_8$
under interaction. This phenomena is known as the interaction
reduced classification of TI/TSC in condensed matter physics.

The interaction reduced classification indicates that eight copies
of the $1d$ TSC is actually in the same phase as the $1d$ trivial
insulator, such that one can (with the help of the interaction)
smoothly tune eight copies of the $1d$ TSC to trivial without
going through a phase transition while respecting the symmetry.
First let us point out that without the interaction the bulk phase
transition can not be avoid. To see this, let us assume
$|\delta|\ll 1$, then the effective Hamiltonian of the $1d$ TSC at
low-energy becomes a $1d$ Dirac fermion $H=\frac{1}{2}\int\dd x\,
\chi^\intercal(\ii\partial_x\sigma^1+m \sigma^2)\chi$ with the
time-reversal symmetry $\dsZ_2^T: \chi\to\mathcal{K}\sigma^3\chi$.
The Dirac mass $m$ is proportional to the parameter $\delta$ in
the lattice model, so that $m>0$ ($m<0$) corresponds to the TSC
(trivial) phase. Tuning from $m>0$ to $m<0$, the bulk gap must
close at $m=0$, which triggers a phase transition. No matter how
many copies of the $1d$ TSC we made, the time-reversal symmetry
will always rule out the additional fermion bilinear mass terms
(which all take the form of $\chi_a^\intercal\ii\sigma^3
A_{ab}\chi_b$ with $A^\intercal=-A$) that anticommute with
$m\sigma^2$, so the bulk phase transition is inevitable at the
free fermion level. The interaction reduced classification means
that the bulk criticality at $m=0$ actually can be removed by the
properly designed interaction.

For $1d$ TSC, this conclusions has been {\it rigorously}
proven\cite{fidkowski1} at the field theory level using the
bosonization approach. The precise conclusion of~\cite{fidkowski1}
is that, for 8 copies of the $1d$ TSC whose low energy field
theory reads $H=\frac{1}{2}\int\dd x\, \sum_{a = 1}^8
\chi^\intercal_a (\ii\partial_x\sigma^1+m \sigma^2)\chi_a$, there
is a $\SO7$ invariant interaction ($\chi_a$ forms a spinor rep of
$\SO7$), which for the entire range of $m$, renders the spectrum
gapped without ground state degeneracy, even for the point where
$\langle \chi^\intercal \sigma^2 \chi \rangle = 0$ (in the field
theory this corresponds to the critical point $m = 0$).

%This means that the field theory $H=\frac{1}{2}\int\dd x\, \sum_{a
%= 1}^8 \chi^\intercal_a (\ii\partial_x\sigma^1 + m
%\sigma^2)\chi_a$ can be driven into a SCSG phase by an $\SO7$
%invariant interaction.

Instead of reproducing the proof in Ref.\,\cite{fidkowski1}, here
we will present a more intuitive argument, which is analogous to
the argument we gave in the main text for the Pati-Salam GUT. The
advantage of this argument is that it can be easily generalized to
any higher spatial dimensions. We first consider two copies of the
$1d$ TSC  coupled to a $\dsZ_2^T$ symmetry-breaking Ising field
$n$, as described by the effective Hamiltonian
$H=\frac{1}{2}\int\dd x\, \chi^\intercal(\ii\partial_x\sigma^{10}
+  m \sigma^{20}+n(x)\sigma^{32} ) \chi$~\footnote{There are too
many possible tuning parameters in the phase diagram if $\dsZ_2^T$
is the only assumed symmetry. Hereinafter we focus on the
particular curve in the phase diagram where the only tuning
parameter is the same mass $m$ for all flavors.}. The strategy is
that we first order the $n$ field to locally gap out the critical
fermions in the bulk at the expense of breaking the $\dsZ_2^T$
symmetry, then we disorder the Ising field $n$ by condensing its
kink defects to restore the symmetry. A fully gapped and
non-degenerate bulk state can be obtained only if the kink is also
fully gapped and non-degenerate. For two copies of the $1d$ TSC,
it is found that the kink of $n(x)$ will trap two Majorana
localized modes, which again defines a complex fermion localized
mode $c$, and the tuning parameter $m$ is coupled to the density
of the complex fermion: $m (c^\dagger c - 1/2)$. Thus at $m = 0$,
the kink has two degenerate states with opposite fermion parity.
Thus the point $m=0$ cannot be driven into a gapped and
nondegenerate state by condensing the kinks. Further analysis
shows that only when we have eight copies of the $1d$ TSC, the
kink, which is a $0d$ object hosting eight Majorana localized
modes, can be trivially gapped out by the interaction for the
entire range of $m$ (following the previous discussion of the $0d$
example). Then when and only when there are $8n$ $1d$ TSC, can we
adiabatically connect $m > 0$ and $m<0$ through condensing the
kinks without closing the gap. And in this case the kink is a
boson that can condense to restore the symmetry. Thus we arrive at
the same conclusion that the $1d$ TSC is $\dsZ_8$ classified under
interaction.

%One explicit way to connect the two states is to start with eight
%copies of the $1d$ TSC and create a segment of the trivial
%insulator inside the TSC, then adiabatically expend the trivial
%segment while shrinking the complemental topological segment by
%moving their domain walls: as the domain wall sweep over the
%entire system, the bulk state will change from the TSC to trivial.
%Since the domain wall can be fully gapped by the interaction, the
%whole processes will not involve any gap closing, and thus a
%smooth path connecting the TSC and the trivial insulator can be
%established.

\subsection{A2. SCSG phase in $1d$: boundary of $2d$ systems}

From the above discussion, we can see there are two equivalent
arguments to demonstrate that a TI/TSC is trivialized by
interaction: (1) the boundary argument by showing that the
boundary of TI/TSC can be driven to the SCSG phase by interaction,
(2) the bulk argument by showing that the bulk
topological-to-trivial phase transition can be removed by
interaction. The study of the interaction reduced classification
of TI/TSC is soon extended to higher spatial dimensions, such as
$2d$ $p\pm\ii p$ TSC \cite{qiz8,yaoz8,zhangz8,levinguz8} (D
class), $3d$ $^3$He-B TSC \cite{chenhe3B,senthilhe3} (DIII class),
$4d$ TSC\cite{xu16}, and higher dimensions TI/TSC in
general\cite{youinversion}. All these examples can be understood
following either the boundary or the bulk arguments concluded
above.

For example, it was shown that the $2d$ $p\pm\ii p$ TSC with a
$\dsZ_2$ symmetry is $\dsZ_8$ classified under interaction. Close
to the topological-to-trivial phase transition, the bulk effective
Hamiltonian in the free fermion limit is
$H=\frac{1}{2}\int\dd^2\vect{x}\chi^\intercal(\ii\partial_1\sigma^{10}+\ii\partial_2\sigma^{30}+m\sigma^{23})\chi$
with the $\dsZ_2:\chi\to\sigma^{03}\chi$ symmetry, where $m>0$
($m<0$) corresponds to the topological (trivial) phase. Following
the boundary argument, the $1d$ boundary of a $2d$ $p\pm\ii p$ TSC
consists of a pair of counter-propagating Majorana edge modes,
described by $H=\frac{1}{2}\int\dd
x\chi^\intercal(\ii\partial_x\sigma^{1})\chi$, which, at the
field-theory level, is the same as the bulk theory of a single
copy of the $1d$ (BDI class) TSC at its $m=0$ critical
point discussed in the last subsection. %with an emergent $\dsZ_2:\chi\to\sigma^1\chi$ symmetry.
As eight copies of the $1d$ TSC can be trivialized by interaction
(due to its $\dsZ_8$ classification) without generating any
fermion bilinear order, it therefore suggests that eight copies of
the $2d$ $p\pm\ii p$ TSC can also be trivialized by interaction,
as its boundary Majorana modes can be driven to the SCSG phase by
the same kind of interaction. This boundary argument can be made
precice\cite{qiz8} by the bosonization formalism as
Ref.\,\cite{fidkowski1}.

For the bulk argument, we can consider two copies of the $2d$
$p\pm\ii p$ TSC close to the $m=0$ critical point while coupling
to an O(2) real boson field $\vect{n}=(n_1,n_2)$, as described by
$H=\frac{1}{2}\int\dd^2\vect{x}\chi^\intercal(\ii\partial_1\sigma^{100}+\ii\partial_2\sigma^{300}+m\sigma^{230}+n_1\sigma^{211}+n_2\sigma^{213})\chi$.
Following the ``defect proliferation argument'', we can first
order the O(2) field $\vect{n}$ to locally gap out the bulk
fermion criticality at the expense of breaking the $\dsZ_2$
symmetry, then we restore the symmetry by proliferating vortices
of $\vect{n}$. It is found that the O(2) vortex will trap two
Majorana localized modes. Then again only for eight copies of the
$2d$ $p\pm\ii p$ TSC, the O(2) vortex will trap eight Majorana
localized modes, which can be gapped out by interaction for the
entire range of $m$, and then condensing the vortices not only
restore the symmetry, but also gives us an adiabatic evolution
from $m>0$ to $m<0$ without closing the bulk gap (one can also
check that the O(2) vortex has a bosonic statistics, thus it is
allowed to condense). Once again, we see that both the boundary
and the bulk arguments lead to the same conclusion that the $2d$
$p\pm\ii p$ TSC is $\dsZ_8$ classified under interaction.

\subsection{A3. SCSG phase in $2d$: boundary of $3d$ systems }

It is soon discovered that the $3d$ $^3$He-B TSC with a
$\mathcal{T}^2=-1$ time-reversal symmetry (DIII class) is
$\dsZ_{16}$ classified. Close to the topological-trivial quantum
phase transition, the bulk effective Hamiltonian in the free
fermion limit is
$H=\frac{1}{2}\int\dd^3\vect{x}\chi^\intercal(\ii\partial_1\sigma^{11}+\ii\partial_2\sigma^{13}+\ii\partial_3\sigma^{30}+m\sigma^{20})\chi$
with the $\dsZ_2^T:\chi\to\mathcal{K}\ii\sigma^{12}\chi$ symmetry,
where $m>0$ ($m<0$) corresponds to the topological (trivial)
phase. The $2d$ boundary of a $3d$ $^3$He-B TSC hosts a gapless
Majorana fermion surface mode, described by
$H=\frac{1}{2}\int\dd^2\vect{x}\,\chi^\intercal(\ii\partial_1\sigma^1+i\partial_2\sigma^3)\chi$,
%which, on the field theory level, only corresponds to half of the
%$2d$ $p\pm\ii p$ TSC at its $m=0$ critical point. The $\dsZ_8$
%classification of the $2d$ $p\pm\ii p$ TSC indicates that if we
%start from eight copies of the $2d$ $p\pm\ii p$ TSC at $m=0$ and
%gradually increase the interaction strength, the system must
%undergo a phase transition from the gapless critical phase (with
%16 Majorana cones at low-energy) to a fully gapped SCSG phase. It
%is numerically verified that the phase transition is
%continuous\cite{doubleQSH}, meaning that there must be a field
%theory describing the transition that gaps out 16 Majorana cones
%by interaction in $2d$, which is then applicable to the surface of
%the $3d$ $^3$He-B TSC.
So if we start with 16 copies of the $3d$ $^3$He-B TSC, the
boundary will host 16 Majorana cones, which can then be driven to
the SCSG phase by interaction. The boundary argument proposed in
Ref.~\cite{senthilhe3,maxhe3} is actually very similar to the bulk
argument in the previous subsection: we can first couple the 16
copies of the $3d$ $^3$He-B TSC to an O(2) vector, and we manually
break the O(2) symmetry by condensing the O(2) vector. Then when
and only when there are 16 copies of $^3$He-B TSC, can the vortex
at the boundary be gapped and nondegenerate and have bosonic
statistics under interaction. Then this means that one can
condense the vortices at the $2d$ boundary to drive the boundary
in to the SCSG phase when and only when the flavor number of the
system is multiple of 16.

To backup the statement by the bulk argument, we may consider four
copies of the $3d$ $^3$He-B TSC near the $m=0$ critical point
while coupling to an O(3) real boson field
$\vect{n}=(n_1,n_2,n_3)$, as described by
$H=\frac{1}{2}\int\dd^3\vect{x}\chi^\intercal(\ii\partial_1\sigma^{1100}+\ii\partial_2\sigma^{1300}+\ii\partial_3\sigma^{3000}+m\sigma^{2000}+n_1\sigma^{1210}+n_2\sigma^{1222}+n_3\sigma^{1230})\chi$.
Again, following the defect proliferation argument, we can first
order the O(3) field $\vect{n}$ to locally gap out the bulk
criticality at the expense of breaking the $\dsZ_2^T$ symmetry,
then we restore the symmetry by condensing O(3) monopoles of
$\vect{n}$. It is found that each O(3) monopole will trap two
Majorana localized modes. So only for 16 copies of the $3d$
$^3$He-B TSC, the O(3) monopole will trap eight Majorana localized
modes and can be therefore trivialized by interaction and safely
condense. In fact, one may also consider disordering the O(3)
field $\vect{n}$ by condensing other topological defects, such as
vortex rings or domain walls. It turns out that\cite{xu16} they
all reach the same conclusion that only 16 copies of the $3d$
$^3$He-B TSC can be trivialized by interaction.

One can see that the same pattern of arguments repeats in every
dimension. The interaction reduced classification of fermionic
TI/TSC states happens in all dimensions, and can be studied
systematically by connecting to the bosonic symmetry protected
topological states\cite{youinversion}.

\section{B. Decomposition and Reconstruction of the Interaction}

In Ref.\,\cite{fidkowski1}, Fidkowski and Kitaev proposed an
$\SO7$ invariant interaction to fully gap out eight local Majorana
zero modes. As quoted in \eqnref{eq: int FK original}, the
interaction Hamiltonian contains 14 four-fermion terms. In this
appendix, we will provide a Hubbard-Stratonovich decomposition of
the Fidkowski-Kitaev (FK) interaction by rewriting the interaction
as inner product of fermion bilinear operators. The decomposition
potentially allows more efficient numerical simulation (for
example the quantum Monte Carlo approach) of the FK interaction in
terms of Yukawa-type interactions.  The decomposition also allow
us to reconstruct many other interactions that has lower symmetry
than $\SO7$ but also gaps out eight local Majorana zero modes with
the same unique ground state. These variant interactions provide
us more choices to gap out the mirror sector fermions, and will be
particularly useful for our purpose of regularizing the GUT on the
lattice. The Yukawa-type interaction also naturally extends to
higher dimensions which provides a general construction of the
interaction that is needed to gap out the gapless fermions in any
dimension.

\subsection{B1. The $0d$ Case: Fidkowski-Kitaev Interaction and its Variants}

Let us start form eight Majorana fermion operators $\chi_i$
($i=1,\cdots,8$) defined by $\{\chi_i,\chi_j\}=2\delta_{ij}$,
which can be pairwise combined into complex (regular) fermion
operators as $f_i=(\chi_{2i-1}+\ii\chi_{2i})/2$ for
$i=1,\cdots,4$. They act on a 16-dimensional Hilbert space, which
admits a set of Fock state basis $\ket{n_1n_2n_3n_4}$ labeled by
the fermion occupation numbers $n_i=f_i^\dagger f_i=0,1$. The FK
interaction is uniquely determined\cite{yoni} by specifying a
reference state $\ket{e_1}$ (the naming convention will be evident
later) in the 16-dimensional Hilbert space, which is also the
ground state to be stabilized by the interaction, \beqn\label{eq:
int FK}
H_\text{FK}=-\sum_{i<j<k<l}V_{ijkl}\,\chi_i\chi_j\chi_k\chi_l,\text{
with }V_{ijkl}=\bra{e_1}\chi_i\chi_j\chi_k\chi_l\ket{e_1}. \eeqn
In this paper, we choose
$\ket{e_1}=(\ket{0000}+\ket{1111})/\sqrt{2}$. The ground state
$\ket{e_1}$ is chosen to be a symmetric state such that it will
not have any fermion bilinear expectation value (not generating
any fermion bilinear mass term which breaks the symmetry in
general),
\beqn\bra{e_1}\chi_i\chi_j\ket{e_1}=\delta_{ij}\quad\text{ for
}i,j=1,\cdots,8.\eeqn It can be explicitly verified that for
$i<j<k<l$, there are 14 non-zero entries of the interaction vertex
tensor $V_{ijkl}$, and all of them take the value of either $+1$
or $-1$, i.e. $V_{ijkl}=\pm1$ if not vanishing. The corresponding
14 four-fermion terms are actually commuting projectors, which
single out their common eigen state $\ket{e_1}$ as the ground
state of $H_\text{FK}$ with an energy $-14$. To see this, we note
that $(\chi_i\chi_j\chi_k\chi_l)^2=1$ so the eigenvalues of
$\chi_i\chi_j\chi_k\chi_l$ are $\pm1$, then $V_{ijkl}=\pm1$
implies that $\ket{e_1}$ is the common eigenstate of every
four-fermion term in $H_\text{FK}$. Moreover, because in general
$\chi_i\chi_j\chi_k\chi_l$ and
$\chi_{i'}\chi_{j'}\chi_{k'}\chi_{l'}$ must either commute or
anticommute with each other (which follows from the Majorana
fermion algebra), but since they have a common eigenstate
$\ket{e_1}$ then they must commute, so the 14 four-fermion terms
are commuting projectors. In the basis that all the projectors are
simultaneously diagonalized (which is also an eigen basis of
$H_\text{FK}$), they must be represented as direct products of
four $\sigma^0$ or $\sigma^3$ matrices,\cite{qiz8} i.e.
$V_{ijkl}\chi_i\chi_j\chi_k\chi_l =\pm\sigma^{abcd}$ (if not
vanishing) where $a,b,c,d=0\text{ or }3$. Any 14 such matrices
$\pm\sigma^{abcd}$ adding together can only produce at most one
eigenstate with eigenvalue $-14$, so we know that $\ket{e_1}$ must
be the unique ground state of $H_\text{FK}$. In conclusion,
$H_\text{FK}$ is a nicely designed interaction that can gap out
eight Majorana zero modes with a non-degenerate ground state
$\ket{e_1}$, on which all the fermion bilinear expectation values
vanish.

In fact, the complete set of eigen basis of $H_\text{FK}$ can be
constructed from the ground state $\ket{e_1}$. Depending on the
fermion parity $F=(-)^{\sum_{i=1}^4n_i}$, they can be divided into
even ($F=+1$) and odd ($F=-1$) parity states, denoted as
$\ket{e_i}$ and $\ket{o_i}$ respectively. \beqn
\ket{e_i}=\chi_1\chi_i\ket{e_1},\quad
\ket{o_i}=\chi_i\ket{e_1}\quad\text{ for }i=1,\cdots,8. \eeqn
$\ket{e_i}$ and $\ket{o_i}$ form a set of orthonormal basis for
the 16-dimensional Hilbert space, on which the FK interaction is
diagonalized \beqn
H_\text{FK}=-14\ket{e_1}\bra{e_1}+2\sum_{i=2}^8\ket{e_i}\bra{e_i}.
\eeqn The orthogonality of the basis follows from $\langle
e_i|e_j\rangle=\langle
o_i|o_j\rangle=\bra{e_1}\chi_i\chi_j\ket{e_1}=\delta_{ij}$ and
$\langle e_i|o_j\rangle=0$ (due to the different fermion parity).
The spectrum of $H_\text{FK}$ can be explicitly verified by acting
\eqnref{eq: int FK} on these basis states. The $\SO7$ symmetry of
the FK interaction\cite{fidkowski1} is  reflected in its spectrum:
the ground state $\ket{e_1}$ is a $\SO7$ scalar, the odd parity
states $\ket{o_i}$ ($i=1,\cdots,8$) form a $\SO7$ spinor, and the
excited even parity states $\ket{e_a}$ ($a=2,\cdots,8$) form a
$\SO7$ vector.

To reveal the $\SO7$ symmetry explicitly, one may introduce the
gamma matrices $\gamma^a$, \beqn\label{eq: gamma
def}(\gamma^a)_{ij}=\ii\bra{e_1}\chi_i\chi_j\ket{e_a}=\ii\bra{e_1}\chi_i\chi_j\chi_1\chi_a\ket{e_1}\quad\text{
for }i,j=1,\cdots,8\text{ and }a=2,\cdots,8.\eeqn With our
specific choice of $\ket{e_1}$, the explicit matrix form of
$\gamma^a$ reads \beqn\label{eq:
gammas}\gamma^{2,\cdots,8}=(\sigma^{002},\sigma^{323},\sigma^{021},\sigma^{203},\sigma^{231},\sigma^{123},\sigma^{211}),\eeqn
which are also the $\gamma^a$ matrices in \eqnref{eq: int 1d
SO(7)}, \eqnref{eq: int 3d SO(7)} and \eqnref{eq: int 0d SO(7)}.
The $\SO7$ generators are then given by $S^{ab}=\chi^\intercal
s^{ab}\chi =\sum_{i,j=1}^8\chi_i (s^{ab})_{ij} \chi_j$ where
$s^{ab}=\frac{1}{2\ii}[\gamma^a,\gamma^b]$ ($a,b=2,\cdots,8$). It
can be checked that $[H_\text{FK},S^{ab}]=0$, so that the FK
interaction has the $\SO7$ symmetry indeed. Using the $\gamma^a$
matrices, the FK interaction can be decomposed as \beqn\label{eq:
int FK HS}
H_\text{FK}=-\frac{1}{4!}\sum_{a=2}^8(\Phi^a\Phi^a-16),\text{ with
}\Phi^a = \chi^\intercal \gamma^a \chi.\eeqn To prove this, we
expand \eqnref{eq: int FK HS} into
$H_\text{FK}=-\frac{1}{4!}\sum_{i,j,k,l}\sum_{a}(\gamma^a)_{ij}(\gamma^a)_{kl}\chi_i\chi_j\chi_k\chi_l
+ \text{const.},$ with some constant energy shift. It can be shown
that
$\sum_{a=2}^8(\gamma^a)_{ij}(\gamma^a)_{kl}=\sum_{a=2}^8\bra{e_1}\chi_i\chi_j\ket{e_a}\bra{e_a}\chi_k\chi_l\ket{e_1}=\bra{e_1}\chi_i\chi_j\chi_k\chi_l\ket{e_1}=V_{ijkl}$
(for $i \neq j \neq k \neq l$), because $\ket{e_{a}}$
($a=2,\cdots,8$)  form a complete set of basis for the two-fermion
exited states, thus $\sum_{a=2}^8\ket{e_a}\bra{e_a}$ is a
resolution identity. So
$H_\text{FK}=-\frac{1}{4!}\sum_{i,j,k,l}V_{ijkl}\chi_i\chi_j\chi_k\chi_l
+\text{const.}=-\sum_{i<j<k<l}V_{ijkl}\chi_i\chi_j\chi_k\chi_l $
matches up with \eqnref{eq: int FK} (and the constant energy shift
can be fixed by considering the cases when $ij$ and $kl$
coincide). Therefore \eqnref{eq: int FK HS} is a
Hubbard-Stratonovich decomposition of the FK interaction. Note
that $\Phi^a$ ($a=2,\cdots,8$) are fermion bilinear operators that
rotates like an $\SO7$ vector, so \eqnref{eq: int FK HS} is
manifestly $\SO7$ invariant. With this decomposition, we can
rewrite the FK interaction in terms of a Yukawa model by
introducing the O(7) real boson field $\phi_a$ such that
$H_\text{FK}=-\sum_{a}\phi_a\Phi^a+\frac{1}{2g}\sum_a\phi_a\phi_a$,
which may allow more efficient numerical simulations by, for
example, the quantum Monte Carlo method.

\begin{figure}[htbp]
\begin{center}
\includegraphics[width=180pt]{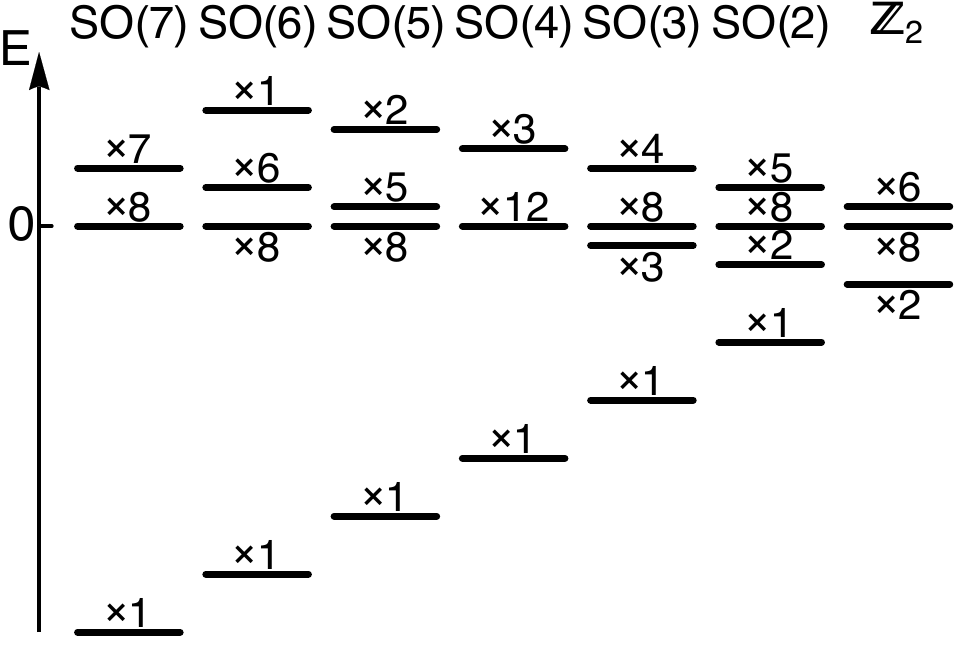}
\caption{Energy spectrum of $\SO{n}$ invariant Yukawa interaction
($n=1$ case labeled by $\dsZ_2$), which is constructed by taking
the last $n$ components of $\Phi^a$ and coupling them to a
$n$-component real boson field.} \label{fig: SO(n)}
\end{center}
\end{figure}

With the fermion bilinear operator $\Phi^a$, we can reconstruct
many other interactions that has a lower symmetry than $\SO7$,
which turns out to be useful for our purpose of designing the
appropriate interaction that has the same symmetry as the GUT that
we try to regularize. To our knowledge, $\SO7$ does not appear as
a gauge group in the mainstream GUT's, so it worth the effort to
explore the variants of the FK interaction with other symmetries.
For example, we can take the last six component of $\Phi^a$
($a=3,\cdots,8$), and construct a $\SO6$ invariant Yukawa
interaction,
$H_\text{int,SO(6)}=-\frac{1}{4!}\sum_{a=3}^8(\Phi^a\Phi^a-16)$,
which is exactly the same interaction as in \eqnref{eq: int fact}
up to some constant energy shift, where
$\Phi^{3,\cdots,8}=\chi^\intercal
\gamma^{3,\dots,8}\chi=f^\intercal(\sigma^{32},-\ii\sigma^{02},\sigma^{20},-\ii\sigma^{23},\sigma^{12},-\ii\sigma^{21})f+\text{H.c.}$
can be read out from \eqnref{eq: gammas} straightforwardly. These
matrices are the same as the $\lambda^a$ matrices defined in
\eqnref{eq: int fact} up to some rearrangement. This $\SO6$
invariant interaction also gaps out eight Majorana zero modes with
a non-degenerated ground state identical to $\ket{e_1}$. To see
this, we start from the representation of the fermion bilinear
operator $\Phi^a$ in the diagonal basis of $H_\text{FK}$,
\beqn\Phi^a=(8\ii\ket{e_1}\bra{e_a}+\text{H.c.})-4\sum_{i,j=1}^8\ket{o_i}(\gamma^a)_{ij}\bra{o_j}.\eeqn
%which follows from the fact that $\Phi^a\ket{e_1}=\sum_{i,j=1}^8(\gamma^a)_{ij}\chi_i\chi_j\ket{e_1}=\sum_{i,j=1}^8\ii\chi_i\chi_j\ket{e_1}\bra{e_1}\chi_i\chi_j\ket{e_a}=-8\ii\sum_{b=2}^7\ket{e_b}\langle{e_b|e_a}\rangle=-8\ii\ket{e_a}$ and that
Then we take the last $n$ components of $\Phi^a$ to construct an $\SO{n}$ invariant Yukawa interaction,
\begin{equation}\label{eq: int SO(n)}
\begin{split}
H_\text{int,SO($n$)}&=-\frac{1}{4!}\sum_{a=8-n+1}^8(\Phi^a\Phi^a-16)\\
&=-2n\ket{e_1}\bra{e_1}+\frac{2n}{3}\sum_{b=2}^{8-n}\ket{e_b}\bra{e_b}+\frac{2(n-4)}{3}\sum_{a=8-n+1}^8\ket{e_a}\bra{e_a},
\end{split}
\end{equation}
whose energy spectrum is plotted in \figref{fig: SO(n)}. As long
as $n\geq 2$, the Majorana zero modes are fully gapped with unique
ground state. From \eqnref{eq: int SO(n)}, we can see the ground
state is always $\ket{e_1}$, identical to the ground state of the
FK interaction, which will not generate any fermion bilinear
expectation value. These conclusions definitely applies to the
$n=6$ $\SO6$ case, which is the interaction that we used to
regularize the Pati-Salam GUT in this paper.

At the first glance, the interaction $H_\text{int}\sim -\sum_a
\Phi^a\Phi^a$ seems to favor the fermion bilinear ordering
$\langle\Phi^a\rangle \neq 0$ at the mean field level, which would
spontaneously break the symmetry (if applying the interaction to a
lattice system), but actually the ordering does not happen.
Because the $\langle\vect{\Phi}\rangle\simeq \vect{\phi}$ ordered
state $\ket{\vect{\phi}}$ (given by the eigen equation
$(\vect{\phi}\cdot\vect{\Phi})\ket{\vect{\phi}}\simeq\ket{\vect{\phi}}$
in the even fermion parity sector, and labelled by the unit vector
$\vect{\phi}$ of the ordering direction) has the wave function
$\ket{\vect{\phi}}=(\ket{e_1}-\ii\sum_{a=2}^8\phi^a\ket{e_a})/\sqrt{2}$,
which is a mixing between the ground state $\ket{e_1}$ and
two-fermion excited states $\ket{e_a}$. Although the state
$\ket{\vect{\phi}}$ indeed gains some interaction energy, but
judging from the energy spectrum given by \eqnref{eq: int SO(n)}
and \figref{fig: SO(n)}, $\ket{e_1}$ will  gain even more energy
than $\ket{\vect{\phi}}$ as long as $n\geq 2$, thus the ordering
does not happen. Physically one may consider $\Phi^a$ as competing
orders that can not make peace with each other, so they compromise
and eventually end up in a quantum superposition state
$\int\dd\vect{\phi}\ket{\vect{\phi}}\simeq\ket{e_1}$ which does
not break the symmetry.

In summary, having specified a (desired) symmetric ground state
$\ket{e_1}$ in the Hilbert space of eight Majorana fermion modes,
we can always find the $\gamma^a$ matrices by \eqnref{eq: gamma
def} and use them to construct the Yukawa interaction
$H_\text{int}\sim-\frac{g}{2}\sum_{a}(\chi^\intercal
\gamma^a\chi)^2$, which, by construction, will single out
$\ket{e_1}$ as its unique ground state. In this construction, the
symmetry group and other details of the interaction term may vary
from one to another, but the flavor number eight for the Majorana
fermions (or four for the complex fermions) always stand out. If
the flavor number is insufficient, the above construction will
cease to work.

\subsection{B2. Higher Dimensions: Generic Yukawa Interaction}

In the above, we have discussed various interactions that can gap
out the Majorana zero modes in the $0d$ system (such as on a
single site or in a monopole core). The construction can be
generalized to higher dimensions to design appropriate
interactions that can remove gapless fermion modes. Following the
defect proliferation argument elaborated in the main text and in
Appendix A, to trivialize a $d$-dimensional gapless fermion system
(if trivializable), we may first couple the fermions to a
symmetry-breaking $\O{d}$ vector order parameter
$\vect{n}=(n_1,\cdots,n_d)$, as
$H=\frac{1}{2}\int\dd^d\vect{x}\chi^\intercal(\ii\partial_\mu\alpha^\mu+n_\mu
\beta^\mu)\chi$, where $\alpha^\mu$ ($\beta^\mu$) (for
$\mu=1,\cdots,d$) are anticommuting symmetric (antisymmetric)
matrices. Then we can condense the order parameter $\vect{n}$ to
gap out the fermions locally, and finally restore the symmetry by
proliferating, say, the monopole defects of the $\vect{n}$ field,
meanwhile the interaction must take effect to remove the fermion
zero modes in the monopole core, such that the monopole can be
safely proliferated. So the interaction in the $d$-dimensional
system must be such designed that it will reduce to the
appropriate interaction (as we discussed previously) in the
monopole core which is capable of gapping out eight Majorana zero
modes. This is our guiding principle to design the interactions in
higher dimensions.

Of cause, one may also consider disordering the O(d) order
parameter $\vect{n}$ by proliferating higher dimensional defects,
such as $1d$ vortex lines or $2d$ domain wall membranes (if they
can be constructed). But as demonstrated in Ref.\,\cite{xu16},
once the monopole proliferation argument goes through, all the
higher dimensional defect proliferation argument will
automatically follow. For example, if we try to proliferate the
vortex lines, we must design the interaction to gap out the $1d$
gapless fermion modes that reside along the vortex line. Then the
problem reduces to its $1d$ version, and we may evoke the defect
proliferate argument again, by considering kink proliferation
along the vortex line, which will then be exactly equivalent to
the monopole proliferation argument. So in the following, we will
only focus on the monopole proliferation argument.

Suppose the monopole configuration is given by $n_\mu\sim x_\mu$
around the monopole core (which has been set to the origin), then
the fermion zero modes $\chi$ in the monopole core are determined
as the common eigenstates of a set of eigen equations:
$\ii\beta^1\alpha^1\chi=\cdots=\ii\beta^d\alpha^d\chi=\chi$. Now
we define a matrix $M=\prod_{\mu=1}^d(\ii\beta^\mu\alpha^\mu)$,
which will act trivially on the monopole modes $M\chi=\chi$ by
construction. So if we consider a fermion bilinear operator
$\Phi^a=\chi^\intercal M\otimes\gamma^a \chi$ in the
$d$-dimensional system, then in the monopole core it will reduce
to $\Phi'^a=\chi^\intercal \gamma^a\chi$ (as $M$ is effectively
set to its eigenvalue $M=1$), which is exactly the operator that
we need to construct the Yukawa interaction in the monopole core.
So the general construction is to start with an $\SO{n}$ invariant
Yukawa interaction in the monopole core
$H_\text{int}=-\frac{g'}{2}\sum_{a=1}^n(\chi^\intercal
\gamma^a\chi)^2$, by reverting the above dimension reduction
procedure, we know that the interaction in the $d$-dimensional
system should be \beqn H_\text{int} =-\frac{g}{2}\sum_{a=1}^n
\Phi^a\Phi^a = -\frac{g}{2}\sum_{a=1}^n(\chi^\intercal
M\otimes\gamma^a\chi)^2,\label{eq: int generic Yukawa}\eeqn in
order to induce the desired interaction in the monopole core. This
is still an $\SO{n}$ invariant local interaction that can act on
each site (or in each unit cell). It shares many similarities with
the FK interaction. For example, its has a fully gapped spectrum
with a non-degenerated ground state $\ket{G}$, whose leading
component is a direct product state of $\ket{e_1}$, i.e.
$\ket{G}\sim\otimes_{\alpha=1}^{m}\ket{e_1}_\alpha$ where $m$ is
the dimension of the matrix $M$, and all the fermion bilinear
expectation values vanish on $\ket{G}$.

To see this, we need to make a few simplifications. Note that the
matrix $M$ is a symmetric matrix by definition, therefore it can
always be diagonalized (by orthogonal transformation) to
$\sigma^3\otimes \mathbf{1}$ whose diagonal elements will be
denoted as $\eta_\alpha=\pm 1$  ($\alpha=1,\cdots,m$) with $m$
being the dimension of $M$. Then the fermion bilinear operator
$\Phi^a=\chi^\intercal M\otimes\gamma^a\chi$ can be decomposed as
a sum of smaller fermion bilinear operators in the $M$ diagonal
basis, i.e. $\Phi^a=\sum_{\alpha=1}^m \eta_\alpha\Phi^a_\alpha$
with $\Phi^a_\alpha=\chi^\intercal_\alpha\gamma^a\chi_\alpha$.
%=(8\ii\ket{e_1}_\alpha{}_\alpha\bra{e_a}+\text{H.c.})-4\sum_{i,j=1}^8\ket{o_i}_\alpha(\gamma^a)_{ij}{}_\alpha\bra{o_j}
So the interaction in \eqnref{eq: int generic Yukawa} can expanded
as $H_\text{int}=-\frac{g}{2}\sum_a(\sum_{\alpha=1}^m\eta_\alpha
\Phi^a_\alpha)^2=-\frac{g}{2}(\sum_{\alpha}\sum_a
(\Phi^a_\alpha)^2+
\sum_{\alpha\neq\beta}\eta_\alpha\eta_\beta\sum_a\Phi^a_\alpha\Phi^a_\beta)$.
The first term $-\frac{g}{2}\sum_{\alpha}\sum_a
(\Phi^a_\alpha)^2=-\frac{g}{2}\sum_\alpha[\sum_a(\chi^\intercal\gamma^a\chi)^2]_\alpha$
is simply the sum of Yukawa interactions over the $\alpha$
sectors, which select out $\ket{e_1}_\alpha$ state as the ground
state in each $\alpha$ sector, so its ground state will be the
direct product of $\ket{e_1}$ states as
$\ket{G_0}=\otimes_{\alpha=1}^m\ket{e_1}_\alpha$. Obviously all
the fermion bilinear expectation values vanish on $\ket{G_0}$. The
second term
$-\frac{g}{2}\sum_{\alpha\neq\beta}\eta_\alpha\eta_\beta
\sum_a\Phi^a_\alpha\Phi^a_\beta$ serves as an off-diagonal
perturbation that mix the $\ket{G_0}$ state with $4k$-fermion
excited states ($k=1,2,\cdots$). Nevertheless the true ground
state $\ket{G}$ of $H_\text{int}$ will still be dominated by
$\ket{G_0}$, as verified by numerics. Also because only
$4k$-fermion excited states are involved in the mixing, so all the
fermion bilinear expectation values will still vanish on
$\ket{G}$, which already implies that $\ket{G}$ is a trivial
representation of the $\SO{n}$ symmetry. To prove that $\ket{G}$
is the unique ground state, we only need to show that the
accidental degeneracy does not occur. To this purpose, we
calculate (by exact diagonalization) the ground state energy $E_0$
(in the even fermion parity sector), the even fermion parity
sector first excited state energy $E_1$, and the odd fermion
parity sector lowest-energy state energy $E_2$ of the interaction
Hamiltonian $H_\text{int}$ in \eqnref{eq: int generic Yukawa}:
\beqn \begin{split}
E_0&=-32g\,m(n+m-1),\\
E_1&=-32g[m(n+m-1)+n-1],\\
E_2&=-32g\,m(n+m-2)+8g(3n-4),
\end{split}\eeqn
which are indeed the three lowest energy levels of $H_\text{int}$.
One can see as long as $n\geq 2$, $E_1$ and $E_2$ never come into
degenerate with $E_0$. Therefore we have shown that $H_\text{int}$
has a fully gapped spectrum with a non-degenerated ground state.
This conclusion applies to all the interaction Hamiltonians that
we constructed in the main text: \eqnref{eq: int 1d SO(7)},
\eqnref{eq: int 3d SO(7)}, \eqnref{eq: int 0d SO(7)}, \eqnref{eq:
int fact}, \eqnref{eq: int 4d SO(6)}, because they all can be
written as \eqnref{eq: int generic Yukawa} (with $n=7$ or $n=6$
and $m$ varies).

\subsection{C. Pati-Salam Model in Majorana Fermion Basis}

In this appendix, we conclude the Pati-Salam model in the Majorana
fermion basis explicitly, such that the relations among the
various symmetry actions and order parameters are clearly exposed.
We first introduce the following 128-component Majorana fermion
field in the $4d$ bulk
\begin{equation}
\chi=\underset{\text{}}{\mat{1\\2}}\otimes\underset{\text{chirality}}{\mat{L\\R}}\otimes\underset{\text{spin}}{\mat{\uparrow\\
\downarrow}}\otimes\underset{\text{flavor}}{\mat{u\\
d}}\otimes\underset{\text{color}}{\mat{r \\ g \\ b \\
w}}\otimes\underset{\text{particle-hole}}{\mat{\Re \psi \\ \Im
\psi}}.
\end{equation}
The layer index $1$ or $2$ labels the fermions that rotate under
$\SU2_1$ or $\SU2_2$ respectively. In the color sector, $r,g,b$
are the three colors of quarks, and $w$ stands for the lepton. In
the particle-hole sector, the complex fermion $\psi$ is written in
terms of two Majorana fermion components as
$\psi=\Re\psi+\ii\Im\psi$. The full effective Hamiltonian in the
$4d$ bulk with the coupling to the O(4) and O(6) fields is given
by $H=H_{\text{TI}}+H_\text{O(4)}+H_{\text{O(6)}}$ (as translated
from \eqnref{4dDirac}, \eqnref{eq: O4} and \eqnref{eq: int 4d
SO(6)}),
\begin{equation}\label{eq: 4dTI in Maj}
\begin{split}
H_{\text{TI}}=\frac{1}{2}\int\dd^4 \vect{x}&\chi^\intercal (\ii\partial_1\sigma^{0310000}-\ii\partial_2\sigma^{0320002}+\ii\partial_3\sigma^{0330000}+\ii\partial_4\sigma^{0100000}+m\sigma^{0200000})\chi,\\
H_\text{O(4)}=\frac{1}{2}\int\dd^4 \vect{x}&\chi^\intercal(-n_0\sigma^{1320001}+n_1\sigma^{1321003}-n_2\sigma^{2322001}+n_3\sigma^{1323003})\chi,\\
H_{\text{O(6)}}=\frac{1}{2}\int\dd^4
\vect{x}&\chi^\intercal(\phi_1\sigma^{0322123}+\phi_2\sigma^{0322203}+\phi_3\sigma^{0322323}+\phi_4\sigma^{3322211}+\phi_5\sigma^{3322021}+\phi_6\sigma^{3322231})\chi.
\end{split}
\end{equation}
Hereinafter $\sigma^{ijk\cdots}=\sigma^i\otimes\sigma^j\otimes\sigma^k\otimes\cdots$ denotes the tensor product of Pauli matrices. The bulk Hamiltonian $H_{\text{TI}}$ has the $\SU4\times\SU2_1\times\SU2_2$ symmetry, given by
\begin{equation}\label{eq: symm in Maj}
\begin{split}
\SU4: &\chi\to e^{\ii\vect{\theta}\cdot\vect{\rho}}\chi,\\
\SU2_1: &\chi\to e^{\ii\vect{\theta}\cdot\vect{\mu}_+}\chi,\\
\SU2_2: &\chi\to e^{\ii\vect{\theta}\cdot\vect{\mu}_-}\chi,\end{split}
\end{equation}
The 15 generators of $\SU4$ are represented as
$\rho^{ij}=\sigma^{p000ijq}$ with $i,j=0,1,2,3$ except for
$ij=00$, while $pq=00$ or $32$ is determined by $ij$ to ensure
that the generator is antisymmetric, \emph{i.e.}
$\rho^{ij\intercal}=-\rho^{ij}$. The generators of $\SU2_1$ and
$\SU2_2$ are represented as
$\vect{\mu}_{\pm}=\frac{1}{2}(\sigma^0\pm\sigma^3)\otimes(\sigma^{001002},\sigma^{002000},\sigma^{003002})$
respectively. The symmetry transformation of the fermion $\chi$
determines the symmetry transformation of the O(4) and O(6)
fields. The O(4) vector $\vec{n}$ rotates under
$\SU2_1\times\SU2_2\simeq\SO4$, and the O(6) vector $\vect{\phi}$
rotates under $\SU4\simeq\SO6$.

The $3d$ boundary of the $4d$ bulk can be considered as the domain
wall of the mass term $m$ flipping across $x_4=0$. The boundary
fermion modes (domain wall fermions) are given by the eigen
equation $\ii\sigma^{0200000}\sigma^{0100000}\chi=\chi$, which
essentially requires to fix $\sigma^{0300000}=1$. Then the
Hamiltonian $H_\text{TI}$ will be reduced to
$H_{3d}=\frac{1}{2}\int\dd^3\vect{x}\chi^\intercal(\ii\partial_1\sigma^{010000}-\ii\partial_2\sigma^{020002}+\ii\partial_3\sigma^{030000})\chi$,
which describes the 16 chiral fermions on the $3d$ boundary.

%\subsection{C1. Localized Fermion Modes in the $\O4$ Monopole Core}

The model Hamiltonian \eqnref{eq: 4dTI in Maj} and symmetry
actions \eqnref{eq: symm in Maj} can be reduced to the O(4)
monopole core. Suppose monopole configuration is described by
$(n_0,n_1,n_2,n_3)\propto(x_1,x_2,x_3,x_4)$ in the vicinity of its
core, then the fermion modes localized in the monopole core are
given by the following eigen equations (which can be derived from
the Schr\"odinger equation\cite{teo}):
\begin{equation}\label{eq: dimred O4}
-\ii\sigma^{1320001}\sigma^{0310000}\chi=
-\ii\sigma^{1321003}\sigma^{0320002}\chi=
-\ii\sigma^{2322001}\sigma^{0330000}\chi=
\ii\sigma^{1323003}\sigma^{0100000}\chi=\chi.
\end{equation}
\eqnref{eq: dimred O4} can be diagonalized to
$\sigma^{3000000}\chi=\sigma^{0300000}\chi=\sigma^{0030000}\chi=\sigma^{0003000}\chi=\chi$
under the following orthogonal transform
\begin{equation}\label{eq: transform O4}
\chi\to
e^{\frac{\ii\pi}{4}\sigma^{2030001}}e^{\frac{\ii\pi}{4}\sigma^{0012000}}e^{\frac{\ii\pi}{4}\sigma^{3202002}}e^{-\frac{\ii\pi}{4}\sigma^{0133002}}\chi,
\end{equation}
which also transforms $\sigma^{0200000}\to-\sigma^{0333002}$,
$\sigma^{0322ij3}\to\sigma^{0330ij3}$,
$\sigma^{3322ij1}\to\sigma^{3330ij1}$. It is straightforward to
see from the diagonalized eigen equations that there are eight
solutions, which corresponds to the eight Majorana modes (or four
complex fermion modes  $f_{1,2,3,4}$ in the main text) localized
in the O(4) monopole core. They can be arranged as
\begin{equation}
\chi=\underset{\text{color}}{\mat{r \\ g \\ b \\
w}}\otimes\underset{\text{particle-hole}}{\mat{\Re f\\ \Im f}}.
\end{equation}
In the subspace of these localized Majorana modes, the model
Hamiltonian is reduced to
\begin{equation}
\begin{split}
H_{\text{TI}}|_{\text{O(4) \ monopole}}&=\chi^\intercal(-m\sigma^{002})\chi,\\
H_{\text{O(6)}}|_{\text{O(4) \ monopole}}&=\chi^\intercal(\phi_1\sigma^{123}+\phi_2\sigma^{203}+\phi_3\sigma^{323}+\phi_4\sigma^{211}+\phi_5\sigma^{021}+\phi_6\sigma^{231})\chi.\\
\end{split}
\end{equation}
The $\SU2_1\times\SU2_2$ symmetry is broken by the O(4) monopole.
The remaining symmetry in the monopole core is the $\SU4$
symmetry, whose generators are reduced to $\rho^{ij}=\sigma^{ijk}$
with $i,j=0,1,2,3$ expect for $ij=00$, and $k=0$ or $2$ determined
by $ij$ to ensure that $\rho^{ij}$ is an antisymmetric matrix. It
is then obvious that the localized fermion modes form a
fundamental representation of the $\SU4$ symmetry.

Judging from the reduced Hamiltonian, these localized fermion
modes will become zero modes at $m=0$ (where the bulk
topological-trivial phase transition is suppose to occur).
However, as discussed in the main text (see \figref{fig:
levels}(c)) and in appendix B, one can construct an $\SO6$
invariant Yukawa interaction
$H_\text{int}=H_{\text{O(6)}}|_{\text{O(4) \
monopole}}+\frac{1}{2g}\vect{\phi}^2$ to gap out the localized
fermion modes for all range of $m$ (including $m=0$). Since the
O(6) Yukawa couplings in the monopole core is originated from the
O(6) Yukawa couplings in the bulk \eqnref{eq: 4dTI in Maj}, so the
corresponding bulk interaction must be given by
$H_\text{int}=H_{\text{O(6)}}+\frac{1}{2g}\int\dd^4\vect{x}\,\vect{\phi}^2$.
After integrating out the bosons field $\vect{\phi}$, one obtains
exactly the interaction we proposed in \eqnref{eq: int 4d SO(6)}
in the main text.

\section{D. Classification of $4d$ Free Fermion Topological Insulators}

The topological insulators/superconductors are classified as the
fermionic symmetry protected topological (FSPT) states. In this
appendix, we fit the various $4d$ topological insulators discussed
in the main text into the ``10-fold way'' classification scheme of
free FSPT states.\cite{ludwigclass1,kitaevclass, ludwigclass2,
wenclass, stoneclass, morimotoclass} In particular, we will focus
on the $\SU4$ and related symmetries, which is important for our
discussion in the main text. In the non-interacting limit, the
classifications are concluded in \tabref{tab: 4d FSPT}. It worth
mention that interaction may further reduce some of the
classifications in the table, as demonstrated in the main text and
reviewed in appendix A. The toy model we discussed corresponds to
the $\U1\times\dsZ_2$ FSPT state, while the Pati-Salam model
corresponds to the $\SU4\times\SU2_1\times\SU2_2$ FSPT state.

\begin{table}[htdp]
\caption{Free fermion classification of some FSPT states in $4d$}
\begin{center}
\begin{tabular}{c|c|cc|c}
Class & Symmetry & Extension Problem & Classifying Space & Classification\\
\hline
\multirow{2}{*}{A} & $\U1$ & \multirow{2}{*}{$\Cl_4\to\Cl_5$} & \multirow{2}{*}{$C_4\cong C_0$} & $\dsZ$ \\
 & $\U1\times\dsZ_2$ & & & $\dsZ\times\dsZ$ \\
\hline
\multirow{2}{*}{C} & $\SU2$ & \multirow{2}{*}{$\Cl_{8,0}\to\Cl_{8,1}$} & \multirow{2}{*}{$R_{-6}\cong R_2$} & $\dsZ_2$\\
 & $\SU2_1\times\SU2_2$ & & & $\dsZ_2\times\dsZ_2$ \\
\hline
AII & $\SU4$ & $\Cl_{10,0}\to\Cl_{10,1}$ & $R_{-8}\cong R_0$ &$\dsZ$ \\
\hline
\multirow{2}{*}{AI} & $\SU4\times\SU2$ & \multirow{2}{*}{$\Cl_{9,3}\to\Cl_{9,4}$} & \multirow{2}{*}{$R_{-4}\cong R_4$} & $\dsZ$ \\
 & $\SU4\times\SU2_1\times\SU2_2$ & & & $\dsZ\times\dsZ$\\
\end{tabular}
\end{center}
\label{tab: 4d FSPT}
\end{table}

Let us start from the $\U1$ FSPT states, which belongs to the
symmetry class A. With the $\U1$ symmetry, the fermion Hamiltonian
can be written in the complex basis as
$H=c^\dagger(\sum_{i=1}^4\ii\partial_i\Gamma^i+m M)c$, where
$\Gamma^i$ ($i=1,\cdots,4$) and $M$ are anti-commuting matrices.
Adding the mass matrix $M$ corresponds to the extension problem
$\Cl_4\to\Cl_5$, whose classifying space is $C_4\cong C_0$, so the
free FSPT classification is given by $\pi_0(C_0)\cong\dsZ$. The
$4d$ $\U1$ FSPT state is also known as the $4d$ quantum Hall (QH)
state. We can stack two $4d$ QH states of opposite chiralities
together to make a non-chiral $4d$ topolotical insulator (TI),
provided an additional $\dsZ_2$ symmetry which acts as the fermion
parity only on one of the chirality. The $\dsZ_2$ symmetry simply
splits the single-particle Hilbert space to two subspaces
(according to the $\pm1$ eigen values of the symmetry operator),
and in each subspace the problem is reduced to the $\U1$ FSPT with
$\dsZ$ classification, so putting together, the $\U1\times\dsZ_2$
free FSPT states are $\dsZ\times\dsZ$ classified in general. The
two $\dsZ$'s stand for the classification of the non-chiral $4d$
TI and that of the chiral $4d$ QH respectively. The
$\U1\times\dsZ_2$ toy model we considered in the main text fits
into the non-chiral $\dsZ$ classification and is hence free from
the perturbative anomaly.

Now we turn to the $\SU2$ FSPT states, which belong to the
symmetry class C. In the Majorana basis, the fermion Hamiltonian
reads $H=\chi^\intercal(\sum_{i=1}^4\ii\partial_i\Gamma^i+m
M)\chi$ where $\Gamma^i$ ($i=1,\cdots,4$) and $M$ are
anti-commuting matrices. For Majorana Hamiltonian, $\Gamma^i$ and
$M$ must also satisfy the symmetry properties that
$\Gamma^{i\intercal}=\Gamma^i$ is symmetric and $M^\intercal=-M$
is anti-symmetric. Denote the $\SU2$ generators as $\mu^a$
($a=1,2,3$), which (in the Majorana basis) are three
anti-commuting and antisymmetric ($\mu^{a\intercal}=-\mu^a$)
matrices. To respect the $\SU2$ symmetry, the Hamiltonian (the
$\Gamma^i$ and $M$ matrices) must commute with these three
generators. All these algebraic relations can be realized in a
single Clifford algebra by embedding the matrices in a larger
space with auxiliary Pauli matrices as
\begin{equation}
\begin{array}{ll}
\sigma^1\otimes\Gamma^i = \alpha^i, & (i=1,\cdots,4)\\
\sigma^2\otimes\mu^a = \alpha^{4+a}, & (a=1,2,3)\\
\sigma^3\otimes 1 = \alpha^8,\\
\sigma^1\otimes M=\beta^1.
\end{array}
\end{equation}
sThen by requiring the symmetric matrices
$\alpha^{p\intercal}=\alpha^{p}$ ($p=1,\cdots,8$) and the
antisymmetric matrices $\beta^{1\intercal}=-\beta^1$ to
anti-commute with each other, all the algebraic properties of
$\Gamma^i$, $M$ and $\mu^a$ are realized. So adding the mass
matrix $M$ corresponds to the extension problem of
$\Cl_{8,0}\to\Cl_{8,1}$, whose classifying space is $R_{-6}\cong
R_2$, therefore the $4d$ $\SU2$ free FSPT classification is given
by $\pi_0(R_2)\cong\dsZ_2$. If the $\SU2$ symmetry is enlarged to
$\SU2_1\times\SU2_2$, the classification will be doubled to
$\dsZ_2\times\dsZ_2$ correspondingly.

Similar classification approach can be applied to the $\SU4$ (and
$\SU4$-related) FSPT states. However, unlike the $\SU2$  group
whose generators are automatically anti-commuting, the 15
generators of the $\SU4$ group do not always anti-commute with
each other. One need to find out the minimal anti-commuting subset
among the 15 generators. It is found that $\Cl_{0,5}\cong\dsC(4)$
is (one of) the minimal Clifford algebra in which the
$\mathfrak{su}(4)$ Lie algebra can be embedded. Denote the
generators of $\Cl_{0,5}$ as $\lambda^a$ ($a=1,\cdots,5$), which
are anti-commuting and antisymmetric
($\lambda^{a\intercal}=-\lambda^a$) matrices, \emph{e.g} a
specific choice may be
$\vect{\lambda}=(\sigma^{102},\sigma^{200},\sigma^{312},\sigma^{320},\sigma^{332})$.
The 15 $\SU4$ group generators can then be obtained either as
$\lambda^a$ or as $\ii\lambda^a\lambda^b$. To respect the $\SU4$
symmetry, it is sufficient to require the Hamiltonian (the
$\Gamma^i$ and $M$ matrices) to commute with $\lambda^a$. All
these algebraic relations can be realized in a single Clifford
algebra by embedding the matrices in a larger space with auxiliary
Pauli matrices as
\begin{equation}
\begin{array}{ll}
\sigma^1\otimes\Gamma^i = \alpha^i, & (i=1,\cdots,4)\\
\sigma^2\otimes\lambda^a = \alpha^{4+a}, & (a=1,\cdots,5)\\
\sigma^3\otimes 1 = \alpha^{10},\\
\sigma^1\otimes M=\beta^1.
\end{array}
\end{equation}
Then by requiring the symmetric matrices
$\alpha^{p\intercal}=\alpha^{p}$ ($p=1,\cdots,10$) and the
antisymmetric matrices $\beta^{1\intercal}=-\beta^1$ to
anti-commute with each other, all the algebraic properties of
$\Gamma^i$, $M$ and $\lambda^a$ are realized. So adding the mass
matrix $M$ corresponds to the extension problem of
$\Cl_{10,0}\to\Cl_{10,1}$, whose classifying space is $R_{-8}\cong
R_0$ (which belongs to the symmetry class AII), therefore the $4d$
$\SU4$ free FSPT classification is given by $\pi_0(R_0)\cong\dsZ$.

The $\SU2$ symmetry can be added to the the $\SU4$ FSPT states,
and the $\dsZ$ classification will not change (but the symmetry
class does change from AII to AI). With the $\SU4\times\SU2$
symmetry, the Clifford algebra embedding scheme can be
\begin{equation}
\begin{array}{ll}
\sigma^1\otimes\Gamma^i = \alpha^i, & (i=1,\cdots,4)\\
\sigma^2\otimes\lambda^a = \alpha^{4+a}, & (a=1,\cdots,5)\\
\sigma^3\otimes\mu^b = \beta^{b}, & (b=1,2,3)\\
\sigma^1\otimes M=\beta^4,
\end{array}
\end{equation}
where the symmetric matrices $\alpha^{p\intercal}=\alpha^{p}$
($p=1,\cdots,9$) and the antisymmetric matrices
$\beta^{q\intercal}=-\beta^q$ ($q=1,\cdots,4$) are anti-commuting
matrices. So adding the mass matrix $M$ corresponds to the
extension problem of $\Cl_{9,3}\to\Cl_{9,4}$, whose classifying
space is $R_{-4}\cong R_4$ (which belongs to the symmetry class
AI), therefore the $4d$ $\SU4\times\SU2$ free FSPT classification
is given by $\pi_0(R_4)\cong\dsZ$. If the symmetry is enlarged to
$\SU4\times\SU2_1\times\SU2_2$, the classification will be doubled
to $\dsZ\times\dsZ$. Again, the two $\dsZ$'s stand for the
classification of the non-chiral $4d$ TI and that of the chiral
$4d$ QH respectively. The Pati-Salam model fits into the
non-chiral $\dsZ$ classification and is hence free from the
perturbative anomaly.

\bibliography{GUT}

\end{document}